\documentclass[preprint]{revtex4-1}
\usepackage{graphicx}
\usepackage{amsmath}
\usepackage{amsthm}
\usepackage{amssymb}
\usepackage{latexsym}
\usepackage{array}
\usepackage{float}
\usepackage{amsfonts}
\usepackage{mathrsfs}
\usepackage{verbatim}
\usepackage{color}
\usepackage[capitalize]{cleveref}
\usepackage{soul}
\usepackage[bb=boondox]{mathalfa}
\usepackage{makecell}
\usepackage{adjustbox,lipsum}

\begin{document}

\title{Estimation of photon number distribution and derivative characteristics of photon-pair sources}

\author{Sang Min Lee}\email{samini@kriss.re.kr}
\affiliation{Korea Research Institute of Standards and Science (KRISS), Daejeon 34113, Rep. of Korea}

\received{\today}
\newcommand{\bra}[1]{\left<#1\right|}
\newcommand{\ket}[1]{\left|#1\right>}
\newcommand{\abs}[1]{\left|#1\right|}
\newcommand{\expt}[1]{\left<#1\right>}
\newcommand{\braket}[2]{\left<{#1}|{#2}\right>}
\newcommand{\commt}[2]{\left[{#1},{#2}\right]}
\newcommand{\tr}[1]{\mbox{Tr}{#1}}
\newcommand{\blue}[1]{\textcolor{blue}{#1}}
\newcommand{\red}[1]{\textcolor{red}{#1}}
\newcommand{\green}[1]{\textcolor{green}{#1}}
\newcommand{\itbf}[1]{\textit{\textbf{#1}}}
\newcommand{\rem}[1]{\red{\st{#1}}}

\begin{abstract}
The evaluation of a photon-pair source employs characteristic metrics like the photon-pair generation rate, heralding efficiency, and second-order correlation function, all of which are determined by the photon number distribution of the source. The photon number distribution, however, can be altered due to spectral or spatial filtering and optical losses, leading to changes in the above characteristics. In this paper, we theoretically describe the effects of different filterings, losses, and noise counts on the photon number distribution and related characteristics. From the theoretical description, an analytic expression for the effective mode number of the joint spectral density is also derived. Compared with previous methods for estimating the photon number distribution and characteristics, an improved methodology is introduced along with a suitable metric of accuracy for estimating the photon number distribution, focusing on photon-pair sources. We discuss the accuracy of the calculated characteristics from the estimated (or reconstructed) photon number distribution through repeated simulations and bootstrapped experimental data.
\end{abstract}

\maketitle

\section{Introduction}\label{s1}

Photon pairs generated via spontaneous parametric down-conversion (SPDC)~\cite{PRL_59_1903} or spontaneous four wave mixing (SFWM)~\cite{NJP_8_67} are primarily used as (heralded) single-photon sources~\cite{OE_24_10733} and entangled photon-pair sources~\cite{PRL_75_4337}, which are the main resources of optical experiments on quantum information processing such as quantum cryptography~\cite{PRL_84_4729}, sensing~\cite{AQT_5_2100164}, and simulation~\cite{NP_8_285}. Many degrees of freedom can be used as information carriers, such as polarization~\cite{PRL_75_4337}, path~\cite{PLA_375_3834}, time-bin~\cite{PRA_66_062308}, spatial mode~\cite{PRL_94_100501} including orbital angular momentum~\cite{N_412_313}, and frequency~\cite{OE_27_1416}. Although frequency entanglement (or correlation) of photon pairs is useful for certain applications (e.g., quantum optical coherence tomography~\cite{QIP_11_903}), in general it is undesirable, especially for multi-photon experiments~\cite{PRL_121_250505}, because of the low spectral purity (indistinguishability or coherence) of single photons. With a few exceptions where group-velocity matching conditions are met~\cite{OE_21_10659} or an aperiodically poled nonlinear crystal is used~\cite{O_5_514}, most photon-pair sources (PPSs) have spectral correlations and use bandpass filters (BPFs) to remove them. However, with BPFs, the main characteristics of the PPS such as (single or coincident) count rate, heralding efficiency, and the value of the second-order correlation function are changed. Typically, the use of BPFs increases photon indistinguishabilities but reduces count rates and heralding efficiencies~\cite{PRA_95_061803}.

In this way, the main characteristics of PPSs change due to the effects of filtering or loss of optical elements frequently used in experiments, but these are fundamentally secondary phenomena caused by changes in the photon number distribution (PND). In this paper, we first theoretically describe the PND of a PPS under ideal circumstances in Sec.~\ref{s2a} and then the changes in the PND under conditions of spectral/spatial filtering and losses in Sec.~\ref{s2b}--\ref{s2d}. In particular, in the process of calculating the probability of two-pair events in Sec.~\ref{s2a}, we derive an analytic expression for the number of effective modes of the joint spectral density, which is the first to our knowledge. In Sec.~\ref{s2e} we describe the characteristics of PPSs, such as pair generation probability, heralding efficiency, and second-order correlation functions, based on the PND. Since the characteristics of PPSs are related to the PND, they are also affected by photon counting errors (noise) in measurement setups. Therefore, in Sec.~\ref{s2f}, we assume the most commonly used measurement settings and describe changes in photon counting results due to noise. Then we analyze the effect of noise on previous methods of estimating the second-order correlation functions using photon counting rates. The results show that, in general, as noise increases, the estimated values of the second-order correlation functions approach 1. It is also discussed that even in the absence of noise, previous methods based on counting rates may generally overestimate the second-order correlation functions for heralded single photons.

Since the characteristics of a PPS are determined by the PND and influenced by noise, accurately estimating the PND by considering (or removing) the influence of noise is equivalent to accurately estimating the characteristics of the PPS. In Sec.~\ref{s3}, we present an improved method for estimating the PND of PPSs that eliminates noise effects and achieves higher accuracy than previous methods. The improved accuracy of our methodology is indirectly confirmed through simulation results for a single-partite PPS, and a qualitative explanation is provided. Additionally, to further clarify the accuracy of the estimated PND, we use a more appropriate metric instead of the previously used fidelity. In the case of a PPS, the probability of no photon is close to 1, so the fidelity between the true and estimated PND is close to 1 no matter how large the differences in other probabilities are. Together with the simulation results of the PND of a PPS, the uncertainties (for 100 repetitions) and errors (from noise) of the estimated values of the second-order correlation functions are also discussed in comparison with previous methods (based on counting rates).

Finally, in Sec.~\ref{s4}, we report and analyze experimental results obtained by applying different combinations of BPFs to a PPS coupled with single-mode fibers (SMFs) based on SPDC. Unlike in simulations, since the true values of the characteristics are not known in experiments, only the uncertainties are evaluated by applying the bootstrapping method to the experimental data, for the reason that this method removes the need for repeated experiments to evaluate uncertainty. For example, in challenging situations where the counting rates are very low, such as PPSs based on ultra-thin materials~\cite{N_613_53}, a single experiment takes a long time and repeat experiments are difficult, making it natural to obtain uncertainties through resampling (bootstrapping) of the experimental data. Detailed theoretical calculation procedures and experimental conditions are covered in the Appendices.

\section{Theoretical description of photon number distribution and derivative characteristics of photon-pair sources}\label{s2}

\subsection{Probabilities of single pair and two pair generation from joint spectral density}\label{s2a}

In previous works \cite{QE_10_1112, NJP_13_033027}, when the signal ($s$) and idler ($i$) photons generated in a SPDC (or SFWM) process were spectrally correlated, the output state of the PPS was shown to be represented by multiple two-mode squeezing operators and second-order correlation function $g^{(2)}$ changes with the number of effective modes as follows:
\begin{align}\label{eq.MTMSV}
\ket \psi 
& = \exp \left( \xi \iint d\omega_s  d\omega_i ~ f(\omega_s,\omega_i) \hat{a}^\dagger _s(\omega_s) \hat a_i ^\dagger(\omega_i) - \rm{ h.c.} \right) \ket0 \notag\\
& = \exp \left( \xi \sum_k c_k  \hat A ^\dagger _k \hat B ^\dagger _k - \rm{ h.c.}  \right) \ket0 
= \bigotimes_k  \exp \left( r_k  \hat A ^\dagger _k \hat B ^\dagger _k - \rm{ h.c.}  \right) \ket0 \notag\\
& = \bigotimes_k \hat S^{(2)}_k (r_k) \ket0.
\end{align}
The normalized joint spectral density (JSD) function $f(\omega_s,\omega_i)$ is Schmidt-decomposed as $\sum_k c_k \theta_k(\omega_s) \phi_k(\omega_i)$ with $\sum_k |c_k|^2 =1$, and effective single-mode operators $\hat A ^\dagger _k$ and $\hat B ^\dagger _k$ are defined as $\int d\omega_s \theta_k(\omega_s) \hat a ^\dagger _s (\omega_s) $ and $\int d\omega_i \phi_k(\omega_i) \hat a^\dagger _i (\omega_i)$, respectively. The operators $\hat X_k$ ($\hat A_k$ or $\hat B_k$) satisfy the canonical commutation relations such as $[\hat X _n,\hat X^\dagger _m]= \hat {\mathbb{1}} \delta_{nm} $ and $[\hat X _n,\hat X _m]=0$. Due to the multi-mode trait, the probability of $j$ photon pairs $p^{(j)}=P(\ket{j,j}_{s,i})$ reveals the difference from the ideal (or single, $k$=1) two-mode squeezed vacuum (TMSV) state, which has a thermal distribution of $p_{TMSV(\xi)}^{(j)}=(1-\mu)\mu^j$ where $\mu=\tanh^2|\xi|\simeq |\xi|^2$ for $|\xi|\ll1$. This change in the photon-pair number distribution leads to a change in $g^{(2)}$ for the $s$ or $i$ mode. When the number of effective modes defined as $K=(\sum_k |c_k|^4)^{-1}$ increases, the distribution $p^{(j)}$ approaches Poissonian and $g^{(2)} =1+K^{-1}$ converges to 1 rather than 2~\cite{NJP_13_033027}.

We now show how to derive the elements of the PND $P_{jk}=P(\ket{j,k}_{s,i})$ analytically from the JSD even in the presence of BPFs. The 1st and 2nd terms of Taylor's expansion of the first line of Eq.~(\ref{eq.MTMSV}) are as follows:
\begin{align}\label{eq.expansion1}
\ket{\psi_1} & =  \left[ \xi \iint d \omega_s d \omega_i f(\omega_s, \omega_i) \hat a ^\dagger  _s (\omega_s) \hat a ^\dagger _i (\omega_i) - {\rm h.c.} \right] \ket 0 \notag \\
& = \xi \iint d \omega_s d \omega_i f(\omega_s, \omega_i) \hat a ^\dagger _s (\omega_s) \hat a ^\dagger _i (\omega_i) \ket 0 ,
\end{align}
\begin{align}
\ket{\psi_2} & = \frac{1}{2}  \left[ \xi \iint d \omega_s d \omega_i f(\omega_s, \omega_i) \hat a ^\dagger  _s (\omega_s) \hat a ^\dagger _i (\omega_i) - {\rm h.c.} \right]^2 \ket 0 \label{eq.expansion2-1}\\
\ket{\psi_{2'}} & = \frac{\xi^2}{2} \iint d \omega_s d \omega_i f(\omega_s, \omega_i) \hat a ^\dagger _s (\omega_s) \hat a ^\dagger _i (\omega_i) \iint d \omega'_s d \omega'_i f(\omega'_s, \omega'_i) \hat a ^\dagger _s (\omega'_s) \hat a ^\dagger _i (\omega'_i)\ket 0 .\label{eq.expansion2-2}
\end{align}
From Eq.~(\ref{eq.expansion1}) and (\ref{eq.expansion2-2}), we can calculate approximated $p^{(1)}$ and $p^{(2)}$ via $\braket{\psi_1}{\psi_1}$ and $\braket{\psi_{2'}}{\psi_{2'}}$, respectively. The difference between $\ket {\psi_2}$ and $\ket {\psi_{2'}}$ in Eq.~(\ref{eq.expansion2-1}) and (\ref{eq.expansion2-2}) is the probability amplitude of $\ket 0$ due to the term $f^*(\omega_s, \omega_i) f(\omega'_s, \omega'_i) \hat a  _s (\omega_s) \hat a  _i (\omega_i) \hat a ^\dagger _s (\omega'_s) \hat a ^\dagger _i (\omega'_i) \ket0$. Since this term corresponds to $p^{(0)}$ rather than to $p^{(2)}$, it is ignored. Similarly, the probability of 1 pair (2 pairs) has components contributed from the higher odd (even) order terms but is negligible due to the magnitude of $|\xi|^{2j}$ ($j\ge3$). The approximated probability of single-pair events, $p^{(1)}$, is calculated as
\begin{align}\label{eq.prob1}
p^{(1)} & \simeq \braket{\psi_1}{\psi_1} \notag \\
 & = |\xi|^2 \bra 0  \iiiint d \omega'_s d \omega'_i d \omega_s d \omega_i ~ f^*(\omega'_s, \omega'_i) f(\omega_s, \omega_i) ~  \hat a _s (\omega'_s) \hat a  _i (\omega'_i) \hat a ^\dagger _s (\omega_s) \hat a ^\dagger _i (\omega_i) \ket 0 \notag \\
& = |\xi|^2 \iint  d\omega_s d \omega_i \left| f(\omega_s, \omega_i) \right|^2  = |\xi|^2,
\end{align}
where $[\hat a (\omega), \hat a ^\dagger (\omega')] = \delta(\omega-\omega')$ and the normalization condition of the JSD is used. Similarly, the approximated probability of two-pair events, $p^{(2)}$, is calculated as
\begin{align}\label{eq.prob2}
p^{(2)} & \simeq \braket{\psi_{2'}}{\psi_{2'}} 
= |\xi|^4 \frac{1}{2}\left(1+\frac{1}{K}\right),
\end{align}
and has a value in $[1/2, 1]~ |\xi|^4$ depending on $K$. In general, in order to obtain the value of $K$, a numerical calculation of the singular value decomposition of the JSD is required~\cite{PRL_84_5304}. Nevertheless, it should be emphasized that we found an {\it analytic expression} of $K$ via the JSD as follows:
\begin{align}\label{eq.KfromJSD}
\frac{1}{K}=\iiiint dx_1 dx_2 dy_1 dy_2 ~ f^*(x_1,y_1) f^*(x_2,y_2) f(x_1,y_2) f(x_2,y_1).
\end{align}
Detailed derivations of Eqs.~(\ref{eq.prob2}) and (\ref{eq.KfromJSD}) are given in Appendix~\ref{asa}. Also, higher-order terms such as $p^{(j)}$ (for $j\geq3$) can be calculated by generalizing the method in Appendix~\ref{asa}, but their magnitudes are on the order of $|\xi|^{2j}$ and they are not required for the PPS characteristics discussed in this paper. We additionally note that, in an ideal situation (without filtering or loss), off-diagonal components such as $P_{jk}$ for $j\neq k$ do not exist. However, these terms can result from (spectral or spatial) filtering effects or optical losses, which is discussed in the next section.

\subsection{Spectral filtering effect on the photon number distribution of a PPS analyzed by JSD segmentation}\label{s2b}

\begin{figure}[ht]
\centering\includegraphics[width=7cm]{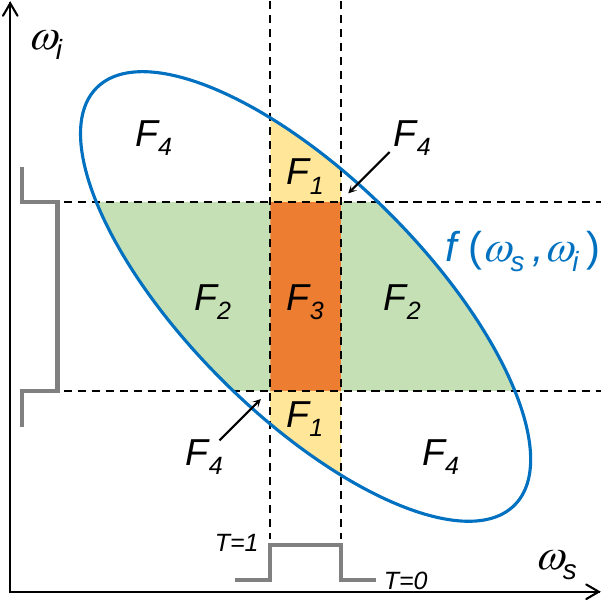}
\caption{Segmentation of the JSD via ideal BPFs.}\label{fig.BPFs}
\end{figure}

Now we consider the case of using BPFs. Assuming ideal BPFs having a transmittance of 1 (0) within (outside) the bandwidth, the JSD is generally divided into four subgroups as shown in Fig.~\ref{fig.BPFs}. The subgroups do not overlap with each other and can be expressed by normalized functions $\{F_j\}$, so that the JSD is re-expressed as $f(\omega_s,\omega_i) = \sum _{j=1}^4 \sqrt{q_j} F_j(\omega_s,\omega_i)$, where $q_j$ is the probability of $F_j$ with respect to $f$. Then the probability of single-pair events $p^{(1)}$ in Eq.~(\ref{eq.prob1}) is divided as $|\xi|^2 (q_1+q_2+q_3+q_4)$, whose contributions are obvious as follows: $q_{1(2)}$ of $F_{1(2)}$ to $P_{10(01)}$ for the single photon probability of the $s~(i)$ mode, $q_3$ of $F_3$ to $P_{11}$ for the photon pair probability between $s$ and $i$ modes, and $q_4$ of $F_4$ to $P_{00}$ for zero photons in the $s$ and $i$ modes (filtered out). Therefore, after BPFs, the photon number distribution matrix, $P$, resulting from single-pair events $p^{(1)}$ is expressed as
\begin{gather}\label{eq.P(1)matrix}
P(p^{(1)}) = P^{(1)} =  |\xi|^2\left(
\begin{tabular} {ccc}
$q_4$&$q_2$&$\cdot$\\
$q_1$&$q_3$&$\cdot$\\
$\cdot$&$\cdot$&$\cdot$
\end{tabular} \right).
\end{gather}

The probability distribution from two-pair events $p^{(2)}$ by JSD segmentation is more complicated to classify. It can be intuitively approximated, though, by combinations of two probabilities ($q_j$, $q_k$) associated with ($F_j$, $F_k$) when $f(\omega_s, \omega_i)$ and $f(\omega'_s, \omega'_i)$ in Eq.~(\ref{eq.expansion2-2}) are replaced by $\Sigma_j \sqrt{q_j} F_j(\omega_s, \omega_i)$ and $\Sigma_k \sqrt{q_k} F_k(\omega'_s, \omega'_i)$, respectively, as
\begin{gather}\label{eq.P(2)matrix_O}
P^{(2)} \sim |\xi|^4\left(
\begin{tabular} {ccc}
$q_4^2$     & $2 q_2 q_4$               &$q_2^2$\\
$2 q_1 q_4$   & $2 q_1 q_2+2 q_3 q_4$     &$2 q_2 q_3$\\
$q_1^2$     & $2 q_1 q_3$               &$q_3^2$
\end{tabular} \right).
\end{gather}
This very rough expression simply considers the number of possible cases and the normalization of probabilities. The exact form of $P^{(2)}$ for the general case (by arbitrary-shaped BPFs) is derived in Appendix~\ref{asb} and given by  Eq.~(\ref{eq.P(2)matrix}). The comprehensive overlap between $F_j$ and $F_k$, which was ignored in Eq.~(\ref{eq.P(2)matrix_O}), is considered in Eq.~(\ref{eq.P(2)matrix}).
 
According to Eqs.~(\ref{eq.P(1)matrix}) and (\ref{eq.P(2)matrix}), when we describe spectrally filtered PPSs, it is reasonable to assume that all components of $P_{jk}$ exist (especially considering together the effects of spatial mode filtering or mismatching and optical losses, which are discussed in following sections). We note that, in general, $P^{(2)}_{jk}$ for $j,k\le 1$ can be neglected because they are smaller than $P^{(1)}_{jk}$ in Eq.~(\ref{eq.P(1)matrix}) by the order of $|\xi|^2$. Also, $P^{(j)}$ for $j \ge 3$ can be calculated by generalization of the method in Appendix~\ref{asb}, but their magnitudes are on the order of $|\xi|^{2j}$ and not necessary for PPS characteristics such as heralding efficiency, pair generation probability, and second-order correlation functions. A more detailed discussion is given in Sec. \ref{s2e}.

\subsection{Spatial filtering effect}\label{s2c}
\begin{figure}[ht]
\centering\includegraphics[width=7cm]{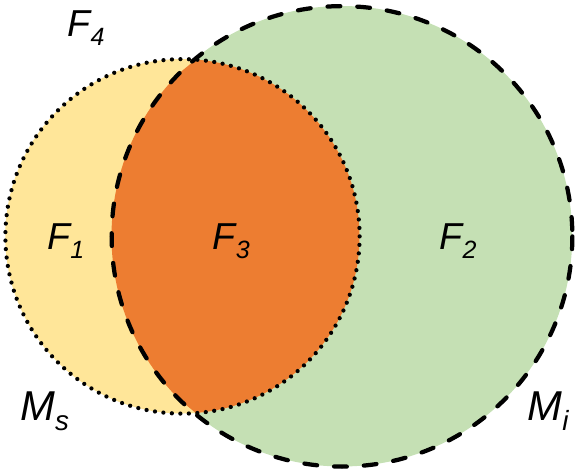}
\caption{Effect of spatial mode mismatch. The dotted (dashed) circle $M_{s(i)}$ indicates the aperture mode used for the $s (i)$ photon.}\label{fig.SPFs}
\end{figure}

The spectral correlation of a PPS constructed by SPDC comes from the energy conservation law ($\omega_p=\omega_s+\omega_i$), which is one part of the phase matching condition (PMC). The other PMC is the momentum conservation law ($\vec k_p=\vec k_s+\vec k_i$), so that $\vec k$ of $s$ and $i$ photons are anticorrelated in the perpendicular plane of $\vec k_p$. Therefore, when we consider the position of the pump beam as the origin and the position of the idler photon moves point-symmetrically with respect to the origin, the positions of the two correlated photons have to coincide within the uncertainty of the PMC~\footnote{More precisely, since momentum is conserved, it is only in the case of nearly degenerate cases $\omega_s \simeq \omega_i$ that the position correlation appears identical. The uncertainty of the correlation is related with the tightness of the PMC, which is mainly determined by the length of nonlinear crystals.}. However, experimentally selected spatial modes for $s$ and $i$ photons via independent apertures or SMF couplings cannot be perfectly matched. Therefore, as shown in Fig.~\ref{fig.SPFs}, the mismatching of two spatial modes of photon pairs produces the same effect (generating the cases of $F_1$ and $F_2$, where a photon exists on only one side) as spectral filtering for frequency-correlated photon pairs, making it hard for the photon number distribution to be ideal (where only $P_{jj}$ exists).
 
\subsection{Optical loss}\label{s2d}

The optics used in PPSs (e.g., lenses, optical fibers, and mirrors used for photon collection, and wave plates for polarization control) have nonideal transmittance (or reflectance) and consequently cause photon losses and degrade the PND. For a single-partite system, the change in PND due to optical loss is expressed as~\cite{JMO_51_1499} 

\begin{align}
P_{(out)} = L(T) \cdot P_{(in)}^{\mathsf T}, \label{eq.singleloss}
\end{align}
where $P=(P_0, P_1, \dots)$ and $L$ is an upper triangular matrix defined by transmittance $T$ for $n, m \ge 0$ as
\begin{align}
L_{nm} = \left\{
\begin{tabular}{lc}
$C^m _n ~ T^n (1-T)^{m-n}$ &  $(m \ge n)$, \\
0                        &  $(m < n)$,
\end{tabular}
\right. \label{Lmatrix}
\end{align}
with the binomial coefficient $C^m_n$. For a bipartite system such as a PPS, the change in PND matrix due to losses is expressed as
\begin{align} \label{eq.LPL}
Q = P_{(out)} = L_s(T_s) \cdot P_{(in)} \cdot L_i^{\mathsf T}(T_i).
\end{align}
Since losses have the effect of shifting the population of $P_{nm}$ to all $P_{n'm'}$ where $n'(m') < n(m)$, it is difficult to maintain the ideal distribution (where only $P_{jj}$ exists) even with small losses.

When photon pairs are not spectrally correlated, the effect of spectral filtering can be considered as a kind of optical loss. As a simple example, suppose $f(\omega_s, \omega_i)$ in Fig.~\ref{fig.BPFs} is a rectangle instead of a tilted ellipse, and the bandwidth ratio of $F_{1(2)}$ to $f$ on the $\omega_{s(i)}$ axis is $T_{s(i)}$ and $R_{s(i)}=1-T_{s(i)}$. When the probabilities of more than three photon pairs are neglected, the photon number distribution according to Eqs.~(\ref{eq.P(1)matrix}) and (\ref{eq.P(2)matrix}) is as follows:
\begin{align*}
P \simeq & ~ P^{(0)} +  P^{(1)} + P^{(2)}  \\
= & ~ \left( 
\begin{tabular}{ccc}
1  & $\cdot$ & $\cdot$ \\
$\cdot$  & $\cdot$ & $\cdot$ \\
$\cdot$  & $\cdot$ & $\cdot$
\end{tabular}
\right)  + 
|\xi|^2 \left( 
\begin{tabular}{ccc}
$R_s R_i$  & $R_s T_i$ & $\cdot$ \\
$T_s R_i$  & $T_s T_i$ & $\cdot$ \\
$\cdot$ & $\cdot$ & $\cdot$
\end{tabular}
\right) + 
|\xi|^4 \left( 
\begin{tabular}{ccc}
$R_s^2 R_i^2$ & $2 R_s^2 T_i R_i$ & $R_s^2 T_i^2$\\
$2 T_s R_s R_i^2$ & $4 T_s R_s T_i R_i$ & $2 T_s R_s T_i^2$  \\
$T_s^2 R_i^2$ & $2T_s^2 T_i R_i$ & $T_s^2 T_i^2$
\end{tabular}
\right) \notag .
\end{align*} 
Actually, the above expression is the same as $ L_s(T_s) \cdot P_{TMSV(|\xi|^2)} \cdot L_i^{\mathsf T}(T_i)$ in Eq.~(\ref{eq.LPL}), where losses ($R_s, R_i$) occurred in the TMSV state. In this simple loss model, any change in losses affects all probability distributions, especially pair probabilities such as $P_{11}$ and $P_{22}$, as described above. On the other hand, consider the case where the ellipse in Fig.~\ref{fig.BPFs} is very thin and long, such that the spectral correlation is very strong. If we assume that the bandwidth of a BPF used for $s~(i)$ photons is very thin (wide), then there will be no $F_1$ region ($q_1=0$). At this time, if the bandwidth of the BPF for the $i$ mode changes slightly, $q_2$ (single photon probability for the $i$ port) is affected but $q_3$ (pair probability) will remain. Accordingly, a comparison of these two opposing cases shows that spectral filtering of a spectrally correlated PPS cannot be described by the simple optical loss model.

\subsection{Characteristics of PPSs from photon number distribution $P_{jk}$}\label{s2e}

Now we discuss the main characteristics of PPSs, namely pair generation probability, heralding efficiency, and (heralded) second-order correlation function, based on the photon number distribution of $P_{jk}$.

\subsubsection{Pair generation probability, $p_g$}\label{s2e1}
Without spectral filtering, the pair generation probability $p_g$ is obviously $p^{(1)}\simeq|\xi|^2$, although $p^{(2)}$ is also related to $K$. For single-pair events after BPFs, the $F_4$ region does not contribute to any counts, nor do $F_1$ and $F_2$ contribute to the coincidence counts. Thus, it is consistent to define $p_g$ as $P_{11}\simeq P^{(1)}_{11}=q_3 |\xi|^2$. Of course, there are also high-order $P^{(j)}_{11}$ ($j\ge2$) components such as $P^{(2)}_{11}$ in Eq.~(\ref{eq.P(2)matrix}), but they are negligible because they are on the order of $|\xi|^{2j}$ ($j\ge2$). In fact, even in the absence of BPFs, components $P_{11}$ caused by $p^{(k)}$ ($k=3,5,\cdots$) exist, but these have already been ignored. As summarized in Table~\ref{tab:prob_pp}, the probabilities of genuine pair events are determined by the segmented JSD via filtering functions, i.e., $F_3 = f(\omega_s,\omega_i) f_s (\omega_s) f_i (\omega_i)/\sqrt{q_3}$, of which the effective mode number is $\kappa_3$.

\begin{table}[h]
\caption{Approximated probabilities of single and two photon pairs with and without BPFs}
\centering
\begin{tabular}{|c||c|c|}
\hline
BPF &$P_{11}$&$P_{22}$ \\ \hline \hline
X & $ |\xi|^2 $    & $ |\xi|^4 (1+1/K)/2$  \\ \hline
O & $ |\xi|^2 q_3$ & $ |\xi|^4 q_3^2 (1+1/\kappa_3)/2$ \\ \hline
\end{tabular}
\label{tab:prob_pp}
\end{table}

\subsubsection{Heralding efficiency and its upper bound, $\eta_H$} \label{s2e2}
The heralding efficiency for $s~(i)$ photons via the Klyshko method~\cite{QE_10_1112} is defined as $\eta_{h,s(i)} \equiv C_{si}/S_{i(s)}$, where $S_{i(s)}$ is the single counting rate for $i~(s)$ photons and $C_{si}$ is the coincidence counting rate between $s$ and $i$ photons. This is not an inherent characteristic of PPSs alone but rather includes the effects of optical losses and the detection efficiencies of the employed single photon detector (SPD). Since a SPD with a finite detection efficiency of $\eta_d$ can be modeled as a beam splitter with transmittance $T=\eta_d$ and an ideal detector~\cite{TIT_26_78}, the counting rates are represented by the photon number distribution after losses (including $\eta_d$), expressed as $Q$ in Eq.~(\ref{eq.LPL}),
\begin{align}
S_s & = r_r \sum_{j=1}^{\infty} \sum_{k=0}^{\infty} Q_{jk} = r_r \sum_{j=1}^{\infty} \left( Q_{j0} + \sum_{k=1}^{\infty} Q_{jk} \right)= r_r \left( Q_{OX} + Q_{OO}\right),\\
S_i & = r_r \sum_{j=0}^{\infty} \sum_{k=1}^{\infty} Q_{jk} = r_r \sum_{k=1}^{\infty} \left( Q_{0k} + \sum_{j=1}^{\infty} Q_{jk}\right)= r_r \left( Q_{XO} + Q_{OO}\right),\\
C_{si} & = r_r \sum_{j=1}^{\infty} \sum_{k=1}^{\infty} Q_{jk} = r_r Q_{OO},
\end{align}
where $r_r$ is the repetition rate of the pump beam, and the subscript $O~(X)$ represents the (non-)detection case of each mode ($Q_{XX}=Q_{00}$). Consequently, $\eta_{h,s}$ and $\eta_{h,i}$ are represented as
\begin{align}
\eta_{h,s} & =\frac{Q_{OO}}{Q_{XO}+Q_{OO}}\le
\eta_{H,s}  \equiv  \frac{P_{OO}}{P_{XO}+P_{OO}} 
\simeq \frac{P^{(1)}_{11}}{P^{(1)}_{01}+P^{(1)}_{11}} = \frac{q_3}{q_2+q_3} , \label{eq.etas}\\
\eta_{h,i} & =\frac{Q_{OO}}{Q_{OX}+Q_{OO}}\le 
\eta_{H,i} \equiv \frac{P_{OO}}{P_{OX}+P_{OO}} \simeq \frac{P^{(1)}_{11}}{P^{(1)}_{10}+P^{(1)}_{11}} = \frac{q_3}{q_1+q_3} . \label{eq.etai}
\end{align}
The equalities and approximations are satisfied for lossless (including $\eta_d=1$) and low pump power ($|\xi|^2 \ll1$) cases, respectively. The upper bound of the heralding efficiency ($\eta_H$) is determined only by the photon number distribution of the PPS, not by extrinsic factors, so it is an inherent characteristic of PPSs. In previous works, simple models of heralding efficiency for PPSs were provided in limiting cases (flat-top filters for a box-shaped JSD~\cite{PRA_92_012329} and Gaussian filters for a tilted Gaussian ellipsoid JSD~\cite{PRA_95_061803}). Our results in Eqs.~(\ref{eq.etas}) and (\ref{eq.etai}) are as intuitive and simple as previous ones while also being applicable to more general cases. As described in Sec.~\ref{s2b} and \ref{s2c}, when a PPS has spectral (spatial) correlation and spectral (spatial) filters are used, even without losses, there are two independent single photon probabilities $q_1$ and $q_2$ for $s$ and $i$ photons, respectively, that do not contribute to coincidence counting. It is therefore natural that $\eta_{H(h),s}$ and $\eta_{H(h),i}$ are independent of each other and not equal. Additionally, there are often cases in which $q_1$ and $q_2$ components are ignored and only paired photons are assumed, and then some reduction factor ($0\le c\le 1$) is introduced, such as $S_s=\tau_s R$, $S_i=\tau_i R$ and $C_{si}=c \, \tau_s \tau_i R$, to match the coincidence counting probability to the single counting probabilities. This, however, can lead to conceptual misunderstandings of both counting rates and the PND of PPSs. Preferably, considering the filtering effects, additional factors should be included: $S_s=\tau_s  (q_1+q_3) R$, $S_i=\tau_i  (q_2+q_3) R$, and $C_{si}=  \tau_s \tau_i \, q_3 R$.

To increase the heralding efficiency of  $s~(i)$ photons, $q_{2(1)}$ has to be reduced. This means that the spectral width of the BPF for the heralded $s~(i)$ mode has to be wide enough that there is no $F_{2(1)}$ in Fig.~\ref{fig.BPFs}, and also that the spatial mode area of the heralded mode, $M_{s(i)}$, should enclose $M_{i(s)}$ for the heralding mode in Fig.~\ref{fig.SPFs}. Therefore, when filtering is used for a correlated degree of freedom, it is difficult for both $\eta_{H,s}$ and $\eta_{H,i}$ to be 1, except for special cases~\footnote{For example, (1) the shape of the JSD is so thin that it can be filtered to have neither $F_1$ nor $F_2$, and (2) multiple separable JSDs are diagonally apart from each other so that only one JSD can be selected through filtering.}. This is because, in general, methods to remove $q_1$ increase $q_2$ in Fig.~\ref{fig.BPFs}, and vice versa.

\subsubsection{Second-order correlation function, $g^{(2)}$} \label{s2e3}
The normalized second-order correlation function shows photon statistics regarding bunching or antibunching~\cite{Loudon}. However, with finite jitter and the resolution of a photon counting system including SPDs and coincidence counting unit (CCU), only time-integrated $g^{(2)}$ is available rather than time-resolved $g^{(2)}(\tau)$ \cite{NJP_13_033027}. First, when BPFs are not used, $g^{(2)}$ of $s$ photons is calculated as follows:
\begin{align}
g^{(2)} & = \frac{\iint dt_1 dt_2 \expt { : \hat n (t_1) ~ \hat n  (t_2) : }  }{
\int dt_1 \expt { \hat n (t_1) } 
~ \int dt_2 \expt { \hat n (t_2) } } 
= \frac{ \expt{: \left( \int d\omega ~ \hat n (\omega) \right)^2  :}   }
{ \expt{ \int d\omega ~ \hat n (\omega)  }^2} 
= \frac{\expt {: \left( \sum_k \hat A_k ^\dagger \hat A_k \right)^2 :} }
{\expt{  \sum_k \hat A_k ^\dagger \hat A_k  }^2} \\
& = \frac{\expt {  \sum_k \hat A_k ^\dagger \hat A_k ^\dagger \hat A_k \hat A_k + 2 \sum_{j<k} \hat n _j \hat n _k } } {\expt{  \sum_k \hat n _k  }^2} 
= \frac{\expt {  \sum_k \left( \hat A_k ^\dagger \hat A_k \hat A_k ^\dagger \hat A_k - \hat A_k ^\dagger \hat A_k \right) + 2 \sum_{j<k} \hat n _j \hat n _k } } 
{\expt{  \sum_k \hat n _k  }^2} \notag \\
& = \frac{\expt {  \sum_k \left( \hat n ^2 _k  - \hat n_k  \right) + 2 \sum_{j<k} \hat n _j \hat n _k } } 
{\expt{  \sum_k \hat n _k  }^2} 
= \frac{\expt {  \sum_k  \hat n ^2 _k  + 2 \sum_{j<k} \hat n _j \hat n _k  } - \expt{ \sum_k \hat n_k} } 
{\expt{  \sum_k \hat n _k  }^2} \notag \\
& = \frac{\expt {  \hat n ^2 _{T} }  - \expt {\hat n_{T} } } 
{\expt{   \hat n _{T}  }^2} 
= \frac{ \sum_n (n^2-n) P_n }  { \left( \sum_n n ~  P_n \right)^2} \simeq  2 \frac{ P_2}{P_1^2}~, \label{eq.g2}
\end{align}
where $\hat n _{T} \equiv \sum_k \hat n_k$ is the total number of $s$ photons without regarding modes $\hat A _k$, and $P_n = P_{nn} \simeq O(|\xi|^{2n})$ is used for the final approximation. In this case, the photon number distributions of the $s$ and $i$ modes are the same, so $g^{(2)}$ of $i$ photons is the same as in Eq.~(\ref{eq.g2}). Using Eqs.~(\ref{eq.prob1}), (\ref{eq.prob2}), and (\ref{eq.g2}), $g^{(2)}$ of $s$ (or $i$) photons is given as $1+(1/K)$, the same as a previous work~\cite{PRL_106_013603}. When BPFs are used for $s$ and $i$ photons, each $g^{(2)}$ can be calculated from each marginal distribution defined as $P_j^{[s]}=P(\ket{j}_s)\equiv \sum_k P_{jk}$ and $P_k^{[i]}=P(\ket{k}_i) \equiv \sum_j P_{jk}$, respectively. However, this is equal to $1+(1/K')$, where $K'$ is the effective mode number of a partial JSD ($\sqrt{q_1}F_1+\sqrt{q_3}F_3$ or $\sqrt{q_2}F_2+\sqrt{q_3}F_3$) considering only the BPF for $s$ or $i$ photons. Therefore, when the bandwidth of the BPF is narrowed ($K'\rightarrow 1$) or widened ($K'\rightarrow K$), $g^{(2)}$ is close to 2 or $1+(1/K)$, respectively.

Now we consider $g^{(2)}$ of heralded photons. Assuming $s$ photons are in the heralded mode, then $g_{h,s}^{(2)}$ can be expressed by converting Eq.~(\ref{eq.g2}) to the conditional probabilities that photons exist in heralding mode $i$ as 
\begin{align}
g_{h,s}^{(2)} & \simeq 2 \frac{ P(n_s=2|n_i \ge 1)}{P^2(n_s=1|n_i\ge1)}
= 2 \frac{ P(n_s=2,n_i\ge1)/P(n_i\ge1)}{P^2(n_s=1,n_i\ge1)/P^2(n_i\ge1)} \notag\\
& = 2 \frac{ P(n_s=2,n_i\ge1) P(n_i\ge1)}{P^2(n_s=1,n_i\ge1)} 
= 2 \frac{ \bigl(  \sum_{k=1} P_{2k} \bigr) \bigl( \sum_{j=0,k=1} P_{jk} \bigr)}{\bigl( \sum_{k=1} P_{1k} \bigr)^2}   \notag\\
& \simeq 2 \frac{ P_{21}+P_{22}  }{ P_{11}}  \frac{ P_{01}+P_{11} }{ P_{11}}
\simeq   2 \frac{ P_{21}+P_{22}  }{ P_{11} ~ \eta_{H,s}} 
\simeq 2 \frac{ q_1 \bigl(1+O_x^{(13)}\bigr) + q_3 \bigl(1+1/\kappa_3 \bigr)/2  }{ \eta_{H,s} } |\xi|^2. 
\label{gh2e}
\end{align}
If there is no filtering and loss, then $P_{01}$ and $P_{21}$ are zero in Eq.~(\ref{gh2e}), and $g_{h,s}^{(2)}$ and $g_{h,i}^{(2)}$ are the same as
\begin{align}
g_h^{(2)} \simeq 2 \frac{P_{22}}{P_{11}} =|\xi|^2 \left(1+\frac{1}{K}\right). \label{gh2i}
\end{align}
 As shown above, since $g_h^{(2)}$ is proportional to $|\xi|^2$, a higher quality heralded single photon source (HSPS) can be obtained at lower pump power in the absence of noise. When a wide (narrow) BPF is used for the $s~(i)$ mode to increase $\eta_{H,s}$, approximations such as $O_x^{(13)}\simeq0$, $\kappa\simeq 1$, $\eta_{H,s} \simeq 1$, and $P^{[s]}_1 \simeq (q_1+q_3) |\xi|^2$ can be used in Eq.~(\ref{gh2e}), so that $g_{h,s}^{(2)}$ is approximated as $2 P_1^{[s]}$. In more detail, since $O_x^{(13)}$ and $\kappa_3^{-1}$ have values in $[0, 1]$, $g_{h,s(i)}^{(2)}$ has a value in the range of $[1,4] \, P^{[s(i)]}_1 / \, \eta_{H,s(i)}$. Therefore, in general, $g_{h}^{(2)}$ increases as $p_g$ increases or $\eta_H$ decreases.

\subsection{Effect of noise on photon counting experiments}\label{s2f}
As discussed in the previous section, the characteristics of PPSs are related to the photon number distribution, and thus they are affected by photon counting errors due to experimental noise. Noise counts are classified into two types based on whether they can be estimated independently with a PPS. Typical examples of type A, independent noise, are dark counts of a SPD and counts due to stray light, while an example of type B, dependent noise, is spontaneous Raman scattering from waveguides in the SFWM process~\cite{OE_12_3737}. Since B-type noise always occurs during the photon pair generation process, it should be regarded as a characteristic of the PPS itself unless there is a method to eliminate or discriminate it. In this section, we examine how noise probabilities affect PPS characteristics, and in the next section, we discuss different means to eliminate its effects in the case of type A. 

\begin{figure}[h]
\centering\includegraphics[width=0.5\textwidth]{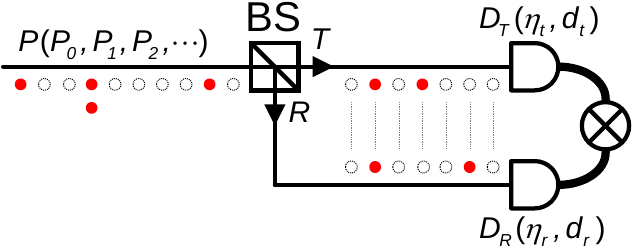}
\caption{Experimental setup of photon counting for a single-partite source, such as the $s$ or $i$ mode of a PPS.}\label{fig.BS_SPDs}
\end{figure}

\subsubsection{Detection probabilities with noise}\label{s2g}
The analysis of noise effects is highly dependent on the experimental setup; here, we assume the most common situation as shown in Fig.~\ref{fig.BS_SPDs}, which is used in our experiments. The port to be measured, i.e., the $s$ or $i$ mode of a PPS, is divided with a beam splitter (BS), and two SPDs are used to count photons, where the transmittance (reflectance) of the BS is $T$~($R=1-T$) and the detection efficiency and noise counting probability of the SPD in the $t$ ($r$) output port are respectively denoted by $\eta_{t(r)}$ and $d_{t(r)}$. The change in PND by the BS can be calculated using Eqs.~(\ref{eq.singleloss}) and (\ref{Lmatrix}). However, since the BS has two output ports, the number of terms to be handled doubles, so it is convenient to simultaneously calculate the detection probabilities of the SPDs. The conversion matrix $M$ from the photon number distribution $P(P_0,P_1,P_2)$ to the detection probabilities is described by a 4$\times$3 matrix as
\begin{align}
 M(T,\eta_t,\eta_r)&
 =\left(
\begin{tabular}{ccc}
1 & $ T(1-\eta_t)+R(1-\eta_r) $     & $ \bigl( T(1-\eta_t)+R(1-\eta_r) \bigr)^2 $ \\
0 & $ R \eta_r $                    & $ R^2 \bigl(1- (1-\eta_r)^2\bigr) + 2T R (1-\eta_t) \eta_r $    \\
0 & $ T \eta_t $                    & $ T^2 \bigl(1- (1-\eta_t)^2\bigr) + 2T R ~ \eta_t (1-\eta_r) $    \\
0 & $ 0 $                           & $ 2 T R ~ \eta_t \eta_r $
\end{tabular}
\right), \label{matrixM}
\end{align}
where the probabilities of more than three photons are ignored. The components of $M \cdot P^{\mathsf T}$ sequentially represent the following four detection cases: ($XX$, $XO$, $OX$, $OO$)$_{tr}$. Additionally, noise events can be modeled as a lower triangular matrix $N$ in a form similar to a previous work~\cite{OE_16_17060} as
\begin{align}
N (d_t,d_r) 
& = \left(
\begin{tabular}{cccc}
$(1-d_t)(1-d_r)$&0   &0   &0\\
$(1-d_t)d_r$    &$1-d_t$&0&0\\
$d_t (1-d_r)$   &0&$1-d_r$&0\\
$d_t d_r$       &$d_t$&$d_r$&1
\end{tabular}
\right), \label{matrixN}
\end{align}
where $d_{t(r)}$ is the noise counting probability of the SPD (including dark counts) at the $t~(r)$ port. This represents increased detection probabilities due to noise counts at the two SPDs. Including optical loss, the final detection probability vector $W$ is expressed as
\begin{align}
W & = N(d_t,d_r) \cdot M(T,\eta_t,\eta_r) \cdot L (\gamma) \cdot P^{\mathsf T} \notag \\
& = N(d_t,d_r) \cdot M(T,\gamma \eta_t,\gamma \eta_r) \cdot P^{\mathsf T}, \label{eq.DP1}
\end{align}
where $L$ can be removed when $\gamma$ is included in $M$. Extending to the case of a bipartite system using two BSs and four SPDs, the detection probability matrix $W$ is expressed as
\begin{align} 
W & =  N_{s}(d_1,d_2) \cdot M_s(T_s, \eta_1, \eta_2 )  \cdot L_s(\gamma_s) \cdot P \cdot L_i^{\mathsf T}(\gamma_i) \cdot M_i^{\mathsf T}(T_i, \eta_3 , \eta_4 ) \cdot N_{i}^{\mathsf T}(d_3,d_4) \notag \\
& =  N_{s}(d_1,d_2) \cdot M_s(T_s, \gamma_s \eta_1 , \gamma_s \eta_2 )   \cdot P \cdot  M_i^{\mathsf T}(T_i, \gamma_i \eta_3 , \gamma_i \eta_4 ) \cdot N_{i}^{\mathsf T}(d_3,d_4),
\label{eq.DP2}
\end{align}
and the 16 detection statuses of $W$ are expressed, in order of $D_1 D_2 D_3 D_4$, as follows:
\begin{align}
\left(
\begin{tabular}{cccc}
$XXXX$    & $XXXO$  & $XXOX$  & $XXOO$\\
$XOXX$    & $XOXO$  & $XOOX$  & $XOOO$\\
$OXXX$    & $OXXO$  & $OXOX$  & $OXOO$\\
$OOXX$    & $OOXO$  & $OOOX$  & $OOOO$ \label{Wcases}
\end{tabular}
\right).
\end{align}

\subsubsection{Noise counting effect on $g^{(2)}$} \label{s2g1}
\begin{figure}[h]
\centering\includegraphics[width=0.5\textwidth]{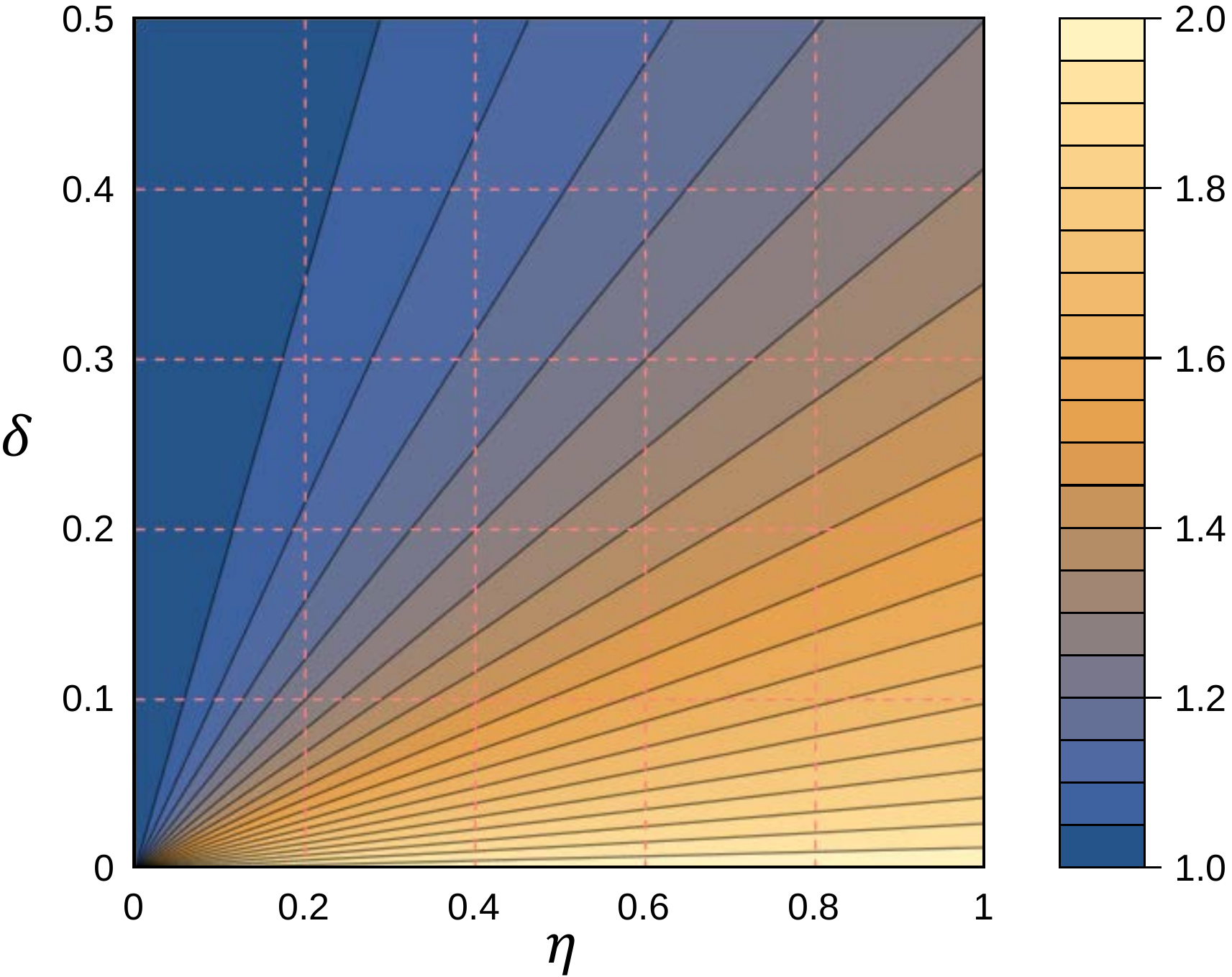}
\caption{Estimated values of $g^{(2)}$ by $C_{tr}/(S_t S_r)$ when the noise counts are comparable to the photon counts ($\eta |\xi'|^2 \sim d = \delta |\xi'|^2$). }\label{fig.g2withN}
\end{figure}

A representative experimental method for estimating $g^{(2)}$ is to calculate $\frac{C_{tr}}{S_t S_r}$~\cite{PRA_78_013844}, where $S_{t(r)}$ is the single-channel counting probability for the $t~(r)$ port and $C_{tr}$ is the coincidence counting probability of both ports in the setup of Fig.~\ref{fig.BS_SPDs}. In the absence of noise and using Eq.~(\ref{matrixM}) with $T=R=1/2$, this is approximated by $g^{(2)}$ in Eq.~(\ref{eq.g2}) as
\begin{align}
\frac{C_{tr}}{S_t S_r} = \frac{ \frac{1}{2} P_2 \eta_t \eta_r}{ \bigl(\frac{1}{2} P_1 \eta_t  + P_2 c_t\bigr)\bigl(\frac{1}{2} P_1 \eta_r  + P_2 c_r\bigr)} \simeq  \frac{2 P_2 }{ P_1^2}, \label{g2CSS}
\end{align}
where $c_{t(r)}$ is some constant. The $P_2$ term of the denominator is ignored since it is smaller than $P_1$ by $|\xi'|^2$. However, in the presence of noise, this approximation is no longer correct. Assuming a more general situation where the noise probability is comparable to $P_1$ as $d_{t(r)}=|\xi'|^2 \delta_{t(r)}$, Eq.~(\ref{g2CSS}) is rewritten as
\begin{align}
\frac{C_{tr}}{S_t S_r} 
& \simeq \frac{d_t d_r + (T \eta_t d_r + R \eta_r d_t  ) P_1 + 2 T R \eta_t \eta_r P_2 }{(d_t + T \eta_t P_1)(d_r + R \eta_r P_1)}
\notag \\
& = \frac{\delta_t \delta_r +  T \eta_t \delta_r + R \eta_r \delta_t    + T R \eta_t \eta_r (1+1/K')
}{ \delta_t \delta_r +  T \eta_t \delta_r + R \eta_r \delta_t    + T R \eta_t \eta_r  } \label{g2CSSappN} ,
\end{align}
where we use $P_1=|\xi'|^2$ and $P_2=|\xi'|^4 (1+1/K')/2$. Additionally, assuming that $T = \frac{1}{2} + \alpha$, $R = \frac{1}{2} - \alpha$, $\eta_t=\eta_r=\eta$, and $\delta_t=\delta_r=\delta$, the above expression becomes
\begin{align} \frac{C_{tr}}{S_t S_r} \simeq \frac{ \delta^2 +  \delta \eta +  (1/4- \alpha^2) \eta^2 (1+1/K') }{  \delta^2 +  \delta \eta + (1/4- \alpha^2) \eta^2 }   \le 1+ \frac{1}{K'} = g^{(2)}   \label{g2ineq} . 
\end{align}
The terms $\delta^2$ and $\delta \eta$ come from the noise counts from both $t$ and $r$ ports and the noise count from one port and the photon count from the other port, respectively. Thus, when the noise is negligible (dark counts $\ll$ photon counts, equivalently $\delta \ll \eta$), the equality in Eq.~(\ref{g2ineq}) is satisfied and a nonideal experimental setup ($\alpha$, $\eta$) has no effect. But when the noise is not negligible, $\frac{C_{tr}}{S_t S_r}$ decreases from $g^{(2)}$ to 1 as $\alpha^2$ or $\delta/\eta$ increases. So we should note that $g^{(2)}$ is no longer robust to loss (included in $\eta$) when the noise counts are comparable to the photon counts. Figure~\ref{fig.g2withN} shows calculated values of $C_{tr}/(S_t S_r)$ in Eq.~(\ref{g2ineq}) according to $\eta$ and $\delta$ when $K'=1$ and $\alpha=0$.

\subsubsection{Noise counting effect on $g_h^{(2)}$} \label{s2g2}
Similar to $g^{(2)}$, $g_h^{(2)}$ is also experimentally estimated through $\frac{C_{tri} ~ S_i}{C_{ti} ~ C_{ri}}$~\cite{JOSAB_24_2972} using the single and coincidence counting probabilities in a setup where the heralded ($s$) mode is divided by a BS, as shown in Fig.~\ref{fig.BS_SPDs}, and the heralding ($i$) mode is directly measured. However, even with approximations of an ideal BS, no noise, and $|\xi|^2\ll1$, $\frac{C_{tri} ~ S_i}{C_{ti} ~ C_{ri}}$ is not the same as $g_h^{(2)}$, depending on the detection efficiency of the heralding mode as below:
\begin{align}
\frac{C_{tri} ~ S_i}{C_{ti} ~ C_{ri}} & \simeq 
\frac{ \left[ \frac{1}{2} P_{21} \eta_t\eta_r \eta_i + \frac{1}{2} P_{22} \eta_t \eta_r \{1-(1-\eta_i)^2\} \right](P_{01}+P_{11})\eta_i}{ \frac{1}{2} P_{11} \eta_t \eta_i ~ \frac{1}{2} P_{11} \eta_r \eta_i} \notag \\
& = 2 \frac{P_{21}+P_{22} (2-\eta_i)}{P_{11}} \frac{P_{01}+P_{11}}{P_{11}} \label{g2hapABC} \\
& \ge 2 \frac{P_{21}+P_{22} }{P_{11}} \frac{P_{01}+P_{11}}{P_{11}} = g_h^{(2)}. \label{gh2ineq}
\end{align}
Therefore, only when $\eta_i \simeq 1$ or $P_{21} \gg P_{22}$ can this be approximated to $g_h^{(2)}$ in Eq.~(\ref{gh2e}). This difference occurs in the $P_{22}$ term because, with a nonideal SPD ($\eta_d < 1$), the detection probability depends on the number of photons. Accordingly, without information about the ratio between $P_{21}$ and $P_{22}$, one cannot correct $\frac{C_{tri} ~ S_i}{C_{ti} ~ C_{ri}}$ to $g_h^{(2)}$ even if $\eta_i$ is known. However, this estimate is limited up to twice $g_h^{(2)}$, so a rough estimate is possible. Meanwhile, noise counts make this situation even worse. For simplicity, if all $d_x$ and $\eta_x$ are equal to $d$ and $\eta$, respectively, then
\begin{align}
\frac{C_{tri} ~ S_i}{C_{ti} ~ C_{ri}} & \simeq 
 2 \frac{  P_{21} + P_{22} (2-\eta) + \frac{P_{11}}{2 T R}  \frac{d}{\eta} }{ P_{11} } ~ \frac{ P_{01} + P_{11} + \frac{d}{\eta} }{ P_{11} }\ge g_h^{(2)}. \label{g2hapABCwithN}
\end{align}
Similar with the case of $g^{(2)}$, the above equation is approximated on the assumption that $d/\eta$ is $O(|\xi|^2)$. However, the estimate increases above $g_h^{(2)}$ as $d$ increases and converges to 1 when $d$ dominates ($d \gg |\xi|^2$). There are two terms that increase the estimate due to noise. The first term, $d/\eta$, is the case when the noise counts at the $i$ port directly increase $S_i$. The second term, $\frac{P_{11}}{2TR}\frac{d}{\eta}$, is the case when a photon pair is measured at the $t~(r)$ and $i$ ports and a noise count simultaneously occurs at the $r~(t)$ port, increasing $C_{tri}$. Although the calculation is performed assuming type A noise, type B also gives the same effect. Therefore, if the noise increases, the characteristics or qualities of the PPS (or HSPS) cannot be accurately estimated. But despite this, by using the inverse matrix of $N$ in Eq.~(\ref{matrixN}), type A noise counts can be removed from the raw counts. With such noise-corrected counts, the noise effects on $g^{(2)}$ and $g_h^{(2)}$ are also corrected. Nevertheless, the difference for the $P_{22}(2-\eta_i)$ term in the numerator of Eq.~(\ref{g2hapABC}) still remains, for the fundamental reason that $P_{21}$ and $P_{22}$ cannot be distinguished effectively because only one detector is used in port $i$. Of course, since this is not a situation that can be solved simply by using two detectors in the $i$ port, an accurate method for estimating $P$ of a PPS is required, which is discussed in the next section.

\section{Improved method for estimating $P$ of a PPS}\label{s3}
In the previous section, considering the realistic situations of spectral/spatial filtering and losses, we explained the reason why the off-diagonal terms of the $P$ matrix for PPSs are nonzero and showed how the main PPS characteristics can be calculated via $P_{jk}$ for $0\le j,k\le2$. We also briefly reviewed commonly used methods for $g^{(2)}$ and $g_h^{(2)}$, and how they may be under- and overestimated due to the noise counts, respectively. Moreover, a discrepancy was found for the case of $g_h^{(2)}$. In this section, we discuss a new methodology for accurately estimating PND.

In order to measure the photon number distribution, of course, a photon number resolving detector (PNRD) is convenient. However, in this paper, it is assumed that on/off SPDs, e.g., a single-photon avalanche photodiode (SPAD) or superconducting nanowire single-photon detector (SNSPD), are used in consideration of practicality and popularity. In previous works~\cite{PRA_70_055801, PRL_95_063602}, the single-partite PND was inferred using an on/off SPD and various attenuators with known parameter values. In this method, the number of events in which no photons were detected was counted in each case of detection efficiency (or attenuation rate), and the original distribution was inferred by solving the extended maximum likelihood (EML) through an expectation--maximization (EM) algorithm. This methodology has been extended to multi-partite systems~\cite{OL_31_3508} and multiple SPDs~\cite{PRA_74_063830}. In the experiments of the above papers, the reported fidelities representing the accuracy of the methodology were over 99~$\%$, but the average photon number of their light sources was greater than or similar to 1. Since the light source considered in the present paper is a PPS of which $p_g$ is much smaller than 1, we simulated whether the above methodology is suitable for this case. Before reviewing the results, we discuss the following points.

First, fidelity defined as $G=\sum_{jk} \sqrt {P_{jk} O_{jk}}$ for two distribution matrices $P$ (true) and $O$ (estimation) is not an appropriate indicator of the estimation accuracy of the PND when $p_g\ll1$. In the case of PPSs, the $P_{00}$ and $O_{00}$ components are very close to 1, so $G$ is obviously close to 1 no matter how different the other components are. In order to fairly compare the size differences for each component, we used the following root mean squared logarithmic error (RMSLE) as an indicator of estimation accuracy:
\begin{align}\label{defRMSLE}
E_{rmsl}(P,O) = \sqrt{ \frac{1}{\mathcal{N}}\sum_{j,k=0} \left( \log_{10}  \frac{P_{jk}+\alpha}{O_{jk}+\alpha} \right)^2 },
\end{align}
where $\alpha$ is a small constant to prevent divergence when the $P$ or $O$ component is zero~\footnote{Normally $\alpha=1$ is used, but in our case we assume that the minimum component ($\sim|\xi|^4$ when either $j$ or $k$ equals 2) is around $10^{-10}$, so we set $\alpha$ to  $10^{-15}$.}, and $\mathcal{N}$ is the total number of elements (size of the matrices, $\sum_{j,k}$). $E_{rmsl}(P,O)$ is independent of the order of $P$ and $O$, and represents the average of the ratios between $P$ and $O$ components. For example, if $P$ and $O$ are identical then $E_{rmsl}=0$, or if all element ratios are $r$ or $1/r$ then $E_{rmsl}=|\log_{10}r|$.

Second, suppose we estimate two single-partite light sources with very small average photon numbers using the previous method. For example, $P'=(\sim1, 10^{-5}, 10^{-10})$ and $P''=(\sim1, 10^{-5}, 10^{-10}/2)$, and their $g^{(2)}$ are 2 and 1, respectively, according to Eq.~(\ref{eq.g2}). After an attenuator with transmittance $T$, each PND is changed to $Q'{}^{(} {}'' {}^{)} (T) = L(T) \cdot P'{}^{(} {}'' {}^{)} {}^{\mathsf T}$ by Eq.~(\ref{eq.singleloss}). Are the experimentally measured frequencies (counts) or detection probabilities of $Q'_0$ and $Q''_0$ different enough to be distinguished?  Due to attenuator effects, the difference between $Q'_0$ and $Q''_0$ is smaller than $10^{-10}/2$, so if the measurement results (or the virtual distributions consisting of repeated trials of measurement counts for $Q'_0$ and $Q''_0$) are indistinguishable, then these two $g^{(2)}$ values are similarly hard to distinguish, regardless of the number of attenuators used. In other words, the $g^{(2)}$ uncertainties of the  reconstructed PND for $P'$ and $P''$ will be much greater than 1. 

\begin{figure}[h]
\centering\includegraphics[width=\textwidth]{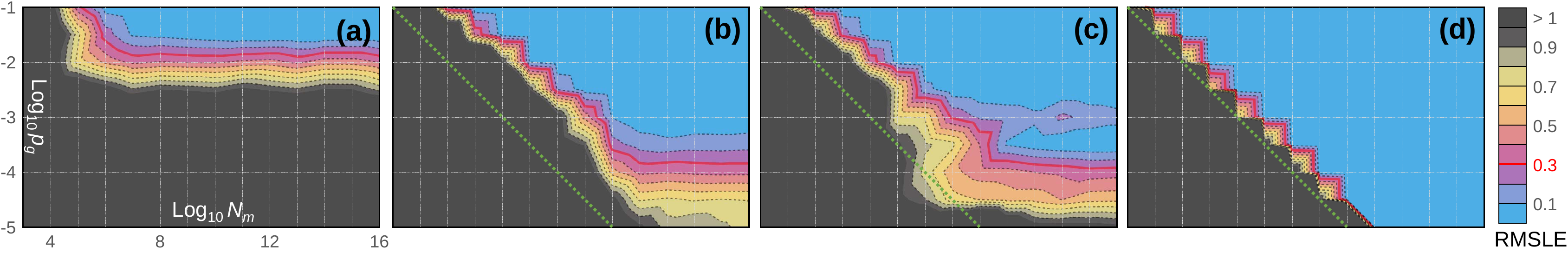}
\caption{Simulation results of $P$ vector estimation (single-partite, $s$ or $i$ photons of a PPS) via the extended maximum likelihood (EML) approach with (a) a single SPD and (b) two SPDs, and via the proposed maximum likelihood (ML) approach using all information including all click events, with (c) a single SPD and (d) two SPDs. Averaged (over 100 times) RMSLE values between $P$ and estimated $P$ according to $N_m$ and $p_g$ are shown.
}\label{fig.SIM1P}
\end{figure}

In simulations, it is assumed that the photon number distribution of the single-partite system ($s$ or $i$ mode of the PPS) has a normalized form of $(\sim1, ~ p_g, ~ p_g^2 g^{(2)} / 2)$, where $g^{(2)}$ is randomly chosen in $[1,2]$ and the number of measurements for each attenuator is $N_m$. Although there are two independent variables, $p_g$ and $g^{(2)}$ (assuming that $P_j$ for $j\ge 3$ is zero), the number of attenuator sets is fixed at 10, which is sufficiently large. Varying $p_g$ and $N_m$ in $[10^{-5}, 10^{-1}]$ and $[10^3, 10^{16}]$, respectively, 100 simulations were conducted for each case, and their RMSLE values were recorded. Figure~\ref{fig.SIM1P} (a) and (b) show the means of the RMSLEs for 100 simulations applying the previous methodology (denoted as EML) using a single SPD (1D) and two SPDs (2D), respectively. Since an RMSLE value of 0.3 means that the average ratio between $P$ and $O$ is about 2 or 1/2, areas with RMSLEs less than 0.3 (blue and violet areas in the figure) are conditions for accurate estimation. As shown in Fig.~\ref{fig.SIM1P} (a), the original scheme (EML-1D) is accurate only in a limited area ($p_g > 10^{-2}$), while $p_g$ of typical PPSs is less than $10^{-3}$. As shown in Fig.~\ref{fig.SIM1P} (b), increasing the number of detectors increases the area of $p_g$ in which accurate estimations can be made. But there is a more effective way, which is to use all the information being measured. EML does not use the information when all detectors are clicked, and this is equivalent to no direct data collection for $P_2$. In other words, since EML-2D directly obtains $P_1$ information from the one-click events, contrary to EML-1D, the size of the region with low RMSLE ($<0.3$) could be increased. Therefore, if we used the information of all click events, the area of accurate estimation can be wider. Figure~\ref{fig.SIM1P} (c) and (d) show the means of the RMSLEs for 100 simulations applying our methodology---ordinary maximum likelihood (ML) using all information including all click events---with a single SPD (1D) and two SPDs (2D), respectively. As shown in Fig.~\ref{fig.SIM1P} (c), the area of accurate estimation when using all information from 1D (ML-1D) is similar to that of EML-2D in Fig.~\ref{fig.SIM1P} (b). Further, by using all information via 2 SPDs (ML-2D), accurate $P$ estimates can be made over a much wider area, as exhibited in Fig.~\ref{fig.SIM1P} (d). Actually, although smaller $p_g$ requires larger $N_m$, there seems to be no limit of $p_g$ (at least up to $10^{-5}$) for accurate estimation of $P_{j\le2}$. Thus, the last method (ML-2D) can be generalized to bipartite systems, ML-2$\oplus$2D, which we apply to the estimation of PPS characteristics in the next section, and can also be easily extended to ML-$j$$\oplus$$k$D if accurate estimates of $P_{jk}$ ($j,k \ge 3$) are required.

\subsection{Methodology via ML using all detection information}\label{s3a}

\begin{figure}[h]
\centering\includegraphics[width=5cm]{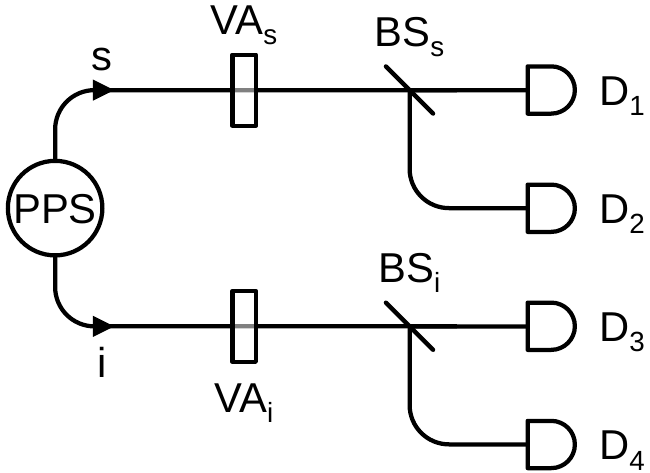}
\caption{Basic scheme of $P$ matrix estimation for a bipartite system. PPS: photon-pair source, $s$: signal, $i$: idler, VA: variable attenuator, BS: beam splitter, D: single-photon detector. }\label{fig.scheme}
\end{figure}

Our scheme is basically similar to that in~\cite{OS_103_90}, as shown in Fig.~\ref{fig.scheme}. Each photon from the PPS passing through a variable attenuator (VA) is divided by a BS to be measured by two SPDs. When the parameters of the VAs, BSs, and SPDs are known, the detection probability matrix is described as Eq.~(\ref{eq.DP2}). Here, $P$ is the matrix of the PND to be estimated, $\gamma_{s(i)}$ is the transmittance of the VA for $s~(i)$ photons, $T_{s(i)}$ is the transmittance of the BS$_{s(i)}$, $\{\eta_j\}$ are the detection efficiencies of the SPDs, and $\{d_j\}$ are the noise probabilities due to dark counts and stray light. We will omit the VAs later, but here assume that $\gamma_{s,i}$ changes according to the $\nu$th setting. For a given set of attenuators $(\gamma_s^{(\nu)},\gamma_i^{(\nu)})$, the $4\times4$ different detection cases in Eq.~(\ref{Wcases}) can be directly counted in an arbitrary time unit. Using these experimentally counted frequencies (or measured counts), $f_{jk}^{(\nu)} \simeq N_m W_{jk}^{(\nu)}$, and the multinomial distribution $W_{jk}^{(\nu)}$, we can construct a log-likelihood function for the $\nu$th case as
\begin{align} \label{eq.LL}
LL_\nu = \sum_{j,k=1} ^{4} f^{(\nu)}_{jk} \log W^{(\nu)} _{jk} .
\end{align}
Since the values of $T_{s,i}$, $\{d_j\}$, $\{\eta_j\}$, and $\{\gamma_{s,i}^{(\nu)}\}$ are known through pre-measurements, $P$ can be estimated by numerically computing the maximum condition of the total log-likelihood function $\mathcal{L}=\sum_{\nu} LL_\nu$.

The main difference between our and the previous method~\cite{PRA_74_063830} is whether or not the probabilities are normalized. In our case, the multinomial distribution $W_{jk}^{(\nu)}$ is normalized as $\sum_{j,k=1}^{4} W_{jk}^{(\nu)}=1$ for each case of $\nu$, so that we can use and construct an ordinary likelihood function. In contrast, the previous method does not utilize all click events, and thus it adopts renormalized probabilities like $W_{jk}^{(\nu)} / \sum_\lambda W_{jk}^{(\lambda)}$ and constructs the extended likelihood function according to the distinct detection cases for $j,k<4$. However, as indirectly shown in Fig.~\ref{fig.SIM1P}, using all click events of $f_{jk}^{(\nu)}$ ($j$ or $k=4$) improves the accuracy of $P$ estimation for a light source with a very small mean photon number. The probabilities $W_{jk}^{(\nu)}$, where either $j$ or $k$ equals 4, are typically as small as $O(|\xi|^{-4})$ because they are derived from $P_2$. Thus, their expected measurement frequencies expressed as $f_{jk}^{(\nu)}\simeq N_m W_{jk}^{(\nu)}$ are also small. If $N_m$ is not large enough such that $f_{jk}^{(\nu)}$ is 0, then the corresponding part $W_{jk}^{(\nu)}$ of the log-likelihood function in Eq.~(\ref{eq.LL}) does not contribute to $P$ estimation, consequently preventing accurate estimation. Thus, $N_m$ has to be sufficiently greater than any $1/W_{jk}^{(\nu)}$. Meanwhile, the presence of VAs corresponding to ($\gamma_s$, $\gamma_i$) generally reduces the size of $W_{jk}$ except for $j=k=1$ (the non-detection case). This means that a low $\gamma$ only impedes the contribution of $W$ by reducing  $f$. Besides, the number of independent variables of $P$ is at most 8, but the number of observations that can be obtained from one setting is 16~\footnote{Not all of these 16 observations are independent of each other. However, if we know all the parameters on the $N$ and $M$ matrices, we can obtain all the information about $P$.}. Thus, we emphasize that when the average photon number is much smaller than 1, our method (ML-2$\oplus$2D) can accurately estimate $P$ of a PPS using sufficiently large $N_m$ without VAs.

\subsection{Simulation results} \label{s3b}

\begin{figure}[h]
\centering\includegraphics[width=\textwidth]{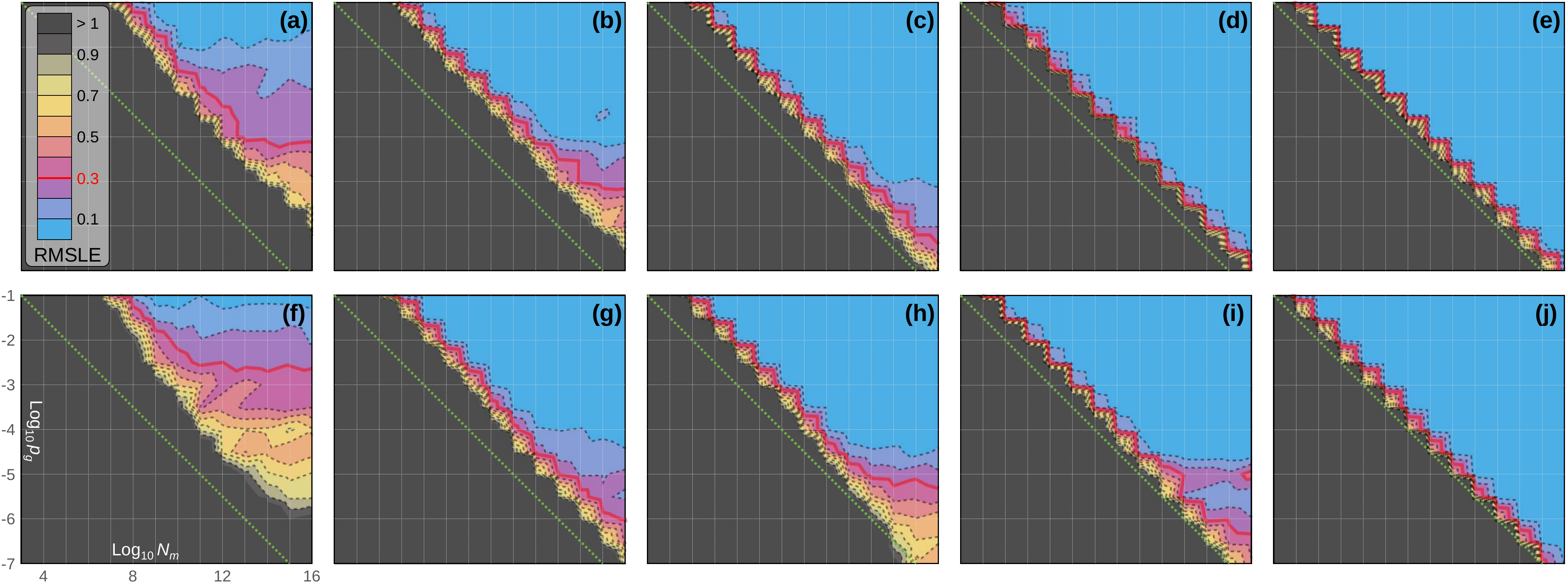}
\caption{Simulation results of $P$ matrix estimation for PPSs via ML using four SPDs (ML-2$\oplus$2D) with SPD detection efficiencies of (a, f) 0.1, (b, g) 0.3, (c, h) 0.5, (d, i) 0.7, and (e, j) 0.9, and attenuator transmittances of (a--e) 1 and (f--j) 0.5 or 1.}\label{fig.SIM2P}
\end{figure}

Figure~\ref{fig.SIM2P} shows the average values of the RMSLEs for 100 simulations according to $p_g$ and $N_m$ when the $P$ matrices of PPSs are estimated via our method. The $P$ matrix in each simulation was randomly chosen as
\begin{align}
P_{jk} = \left\{ 
\begin{array}{ll}
p_g ~ r & \mbox{    if ~ $\max(j,k)=1$},\\
p_g^2 ~ r & \mbox{    if ~ $\max(j,k)=2$},\\
\end{array}
\right.
\end{align}
where $r$ is a random real number in [0.5, 1.5], $P_{00}$ is set to $\sum_{j,k=0}^2 P_{jk}=1$, and no noise is assumed as $\{d_j\}=0$. The detection efficiencies of all SPDs were set as 0.1, 0.3, 0.5, 0.7, and 0.9 for the plots in (a, f), (b, g), (c, h), (d, i), and (e, j), respectively. The number of VA sets used in the upper (a--e) and lower (f--j) cases is 1 and 4, respectively. In fact, cases (a--e) are equivalent to no VAs, i.e., $\gamma_{s,i}=1$. On the other hand, in cases (f--j), $\gamma_{s,i}$ were independently selected as 0.5 or 1, so the total number of measurement resources used in the lower cases was 4 times greater than that in the upper cases. In the lower cases, despite using more resources at the same detection efficiency, the estimation accuracy is somewhat degraded rather than improved due to the reduced $f_{jk}^{(\nu)}$ and $W_{jk}^{(\nu)}$ by losses ($\gamma_{s,i}^{(\nu)} <1$). This clearly shows that VAs are not necessary to estimate $P$ of PPSs. Likewise, since $W_{jk}$ also decreases as $\eta_d$ decreases, more accurate $P$ estimation is possible with higher $\eta_d$ of the SPDs. In particular, from the simulation results in Fig.~\ref{fig.SIM2P} (c), we can cautiously conclude that our method accurately estimates $P$ of PPSs under typical experimental conditions ($\eta_d \ge 0.5$ and $p_g \ge 10^{-5}$).

\subsection{Estimation accuracy of $g^{(2)}$ and $g_h^{(2)}$ via reconstructed $P$} \label{s3c}

\begin{figure}[h]
\centering\includegraphics[width=0.4 \textwidth]{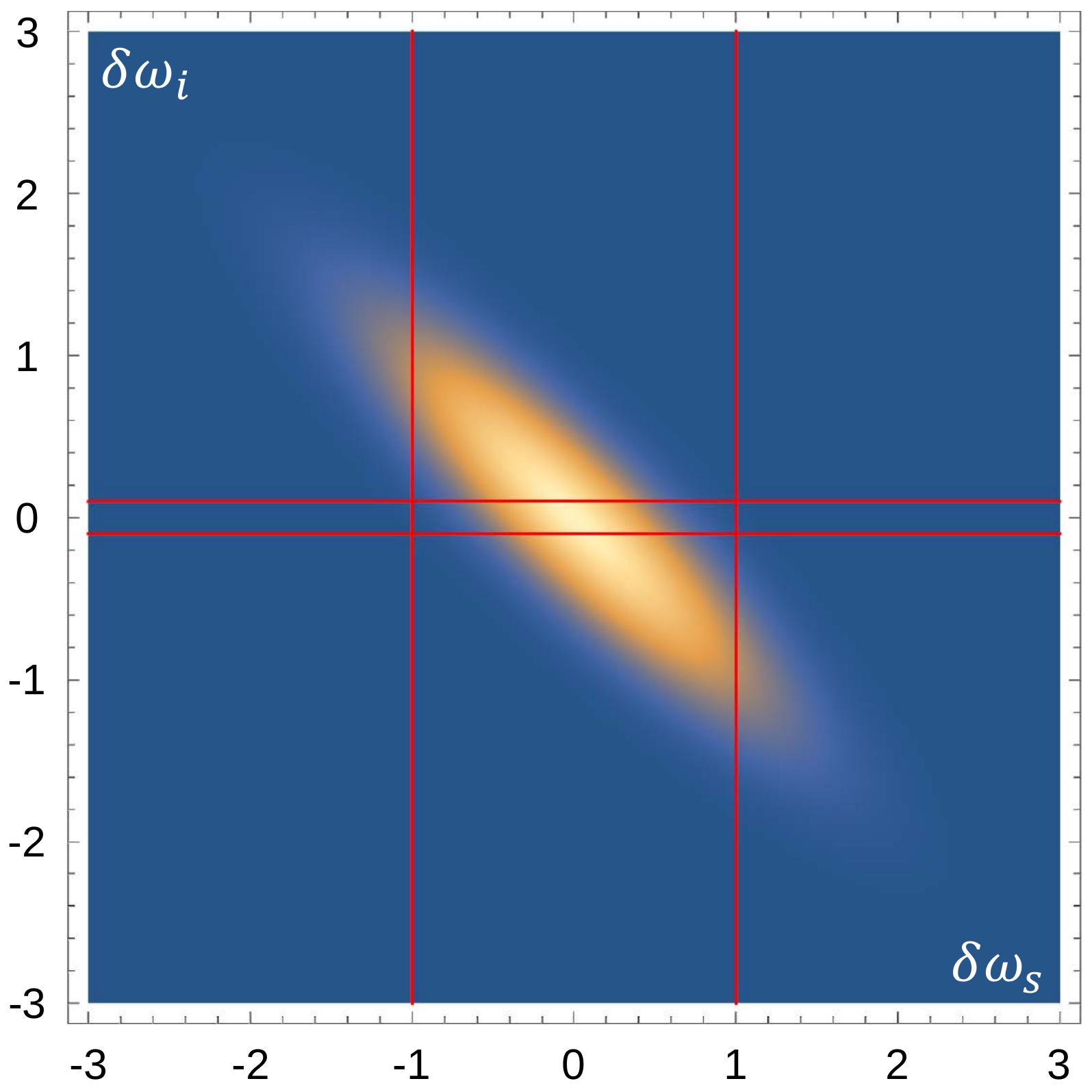}
\caption{Joint spectral density function and widths of ideal BPFs ($s$: 2, $i$: 0.2) used for simulation. }\label{fig.SIM_JS_FF}
\end{figure}

\begin{figure}[h]
\centering\includegraphics[width=0.8\textwidth]{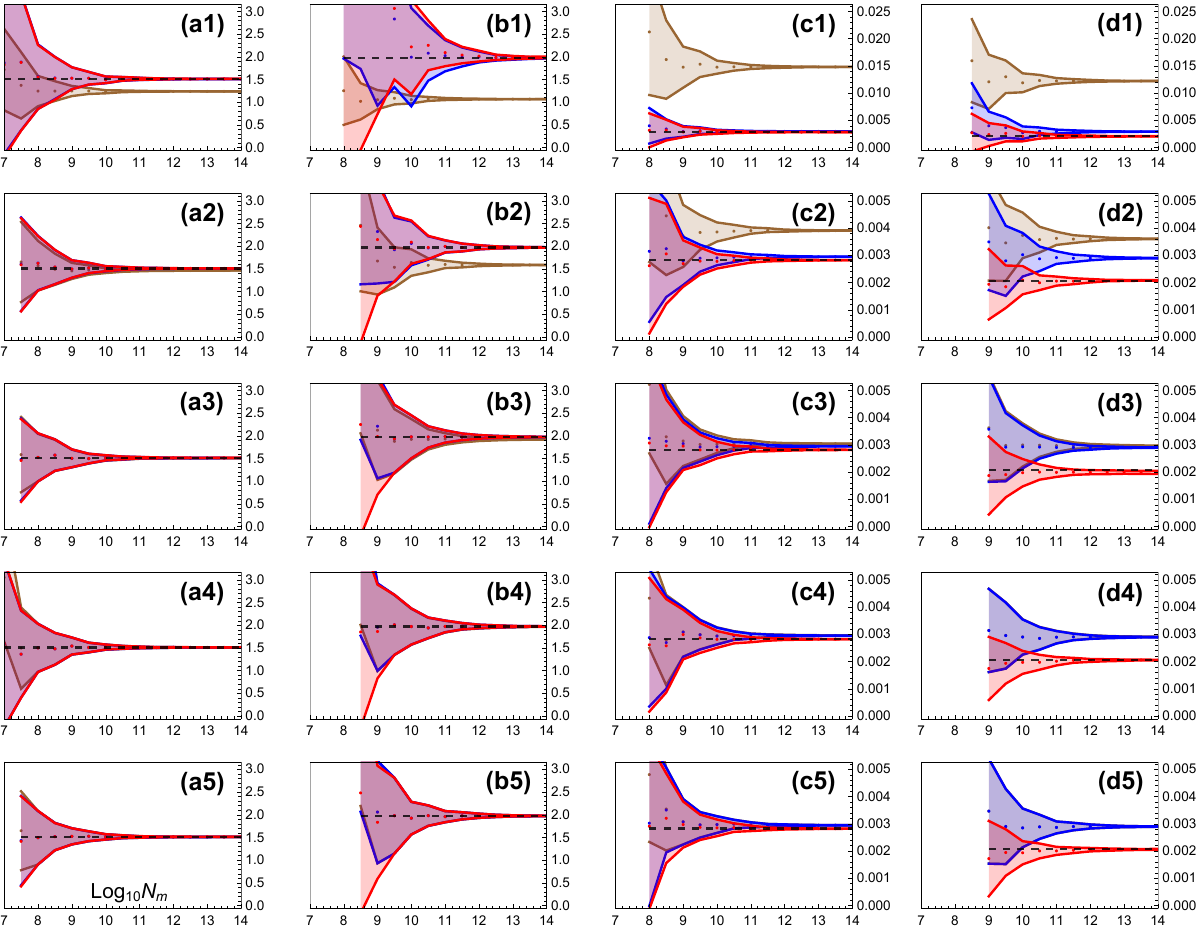}
\caption{Estimated (a, b) $g^{(2)}$ and (c, d) $g_h^{(2)}$ for ($s$, $i$) modes, respectively, with different noise conditions (distinguished by the number following the letter) according to different methodologies: raw counts (brown), noise-corrected counts (blue), and reconstructed $P$ via ML-2$\oplus$2D (red).}\label{fig.SIMg2}
\end{figure}

The reason for estimating $P$ is to analyze the characteristics of a PPS, as described in Sec.~\ref{s2e}. In addition, the accuracy of $g^{(2)}$ and $g_h^{(2)}$ calculated from the reconstructed $P$ is also our interest~\footnote{Since $p_g$ and $\eta_H$ derived from $P^{(1)}$ can be inferred much more accurately than $g^{(2)}$ and $g_h^{(2)}$ derived from $P^{(2)}$, we do not discuss them here but show the related experimental results in Sec.~\ref{s4}.}, so we investigated a typical case as shown in Fig.~\ref{fig.SIM_JS_FF}. This is a configuration that uses a combination of wide (for the heralded mode, $s$) and narrow (for the heralding mode, $i$) BPFs to generate a HSPS with high $\eta_H$. We also assumed typical parameters of $|\xi|^2 = 10^{-3}$, $\eta_d = 0.5$, and optical losses ($T=0.9$) from Eq.~(\ref{eq.LPL}) at both $s$ and $i$ ports for a realistic simulation. In order to check the effect of noise (dark) counts, $d$ was set from $10^{-4}$ to $10^{-8}$, which corresponds to $10^4$ to 1 cps when the repetition rate of the pump laser is about 100 MHz. In Fig.~\ref{fig.SIMg2}, the points and areas colored in brown, blue, and red represent the means and standard deviations of $g^{(2)}$ or $g_h^{(2)}$ values calculated from 100 simulations of raw counts, noise-corrected counts, and reconstructed $P$ (via ML-2$\oplus$2D), respectively. Figure~\ref{fig.SIMg2} (a, b) plot the results of estimated $g^{(2)}$ and (c, d) plot that of $g_h^{(2)}$ for the ($s$, $i$) modes. Graphs in the same row are the results under different noise conditions, specifically $d$ of $10^{-4}$ (first row), $10^{-5}$ (second), $10^{-6}$ (third), $10^{-7}$ (fourth), and $10^{-8}$ (fifth). The dashed black lines are the true values of $g^{(2)}$ and $g_h^{(2)}$. The noise-corrected counts were obtained by removing the increments due to noise counts using the inverse matrix of $N$ in Eq.~(\ref{matrixN})~\footnote{After removing the noise effect, counts less than 0 may occur, in which case these are treated as 0.}. As described in Eq.~(\ref{g2ineq}) and Eq.~(\ref{gh2ineq}), the estimated $g^{(2)}$ and $g_h^{(2)}$ via raw counts (including noise counts) are lower and higher than the true values, respectively, and converge to the results of the noise-corrected counts (in blue) as $d$ decreases. Additionally, the estimated $g^{(2)}$ from noise-corrected counts is as accurate as that from reconstructed $P$, while $g_h^{(2)}$ is not. If $P_{21} \gg P_{22}$ in Eq.~(\ref{g2hapABC}), $g_h^{(2)}$ could be approximated using the noise-corrected counts, which corresponds to the case of the $s$ mode, because $P_{21}$ and $P_{22}$ are approximately $q_1 q_3$ and $q_3^2$ in Eq.~(\ref{eq.P(2)matrix_O}), respectively, and $q_1 > q_3$ due to the configuration of the BPFs in Fig.~\ref{fig.SIM_JS_FF}. Thus, in Fig.~\ref{fig.SIMg2} (c), the results in blue and red are very similar but slightly distinguished at large $N_m$. On the other hand, the $i$ mode corresponds to the opposite case ($q_2 q_3 \sim P_{12} \ll P_{22} \sim q_3^2$ where $q_2 \simeq 0$), creating a noticeable discrepancy between the estimated and true values of $g_h^{(2)}$. More specifically, since $\eta_d=0.5$ and $P_{12} \ll P_{22}$, $g_h^{(2)}$ is overestimated by about a factor of $1.5=(2-\eta_d)$, which is also consistent with the simulation results (convergence values of blue and red: $\sim 0.003$ and $\sim 0.002$) in Fig.~\ref{fig.SIMg2} (d). Meanwhile, the estimates via reconstructed $P$ are accurate regardless of noise and have similar uncertainties as the previous methods. We note that the magnitudes of uncertainties according to $N_m$ are different depending on the mode, because the widths of the two BPFs for the $s$ and $i$ modes are asymmetric (2 : 0.2), resulting in a difference in the number of detection counts.

\section{Experiments}\label{s4}

\begin{figure}[h]
\centering\includegraphics[width=\textwidth]{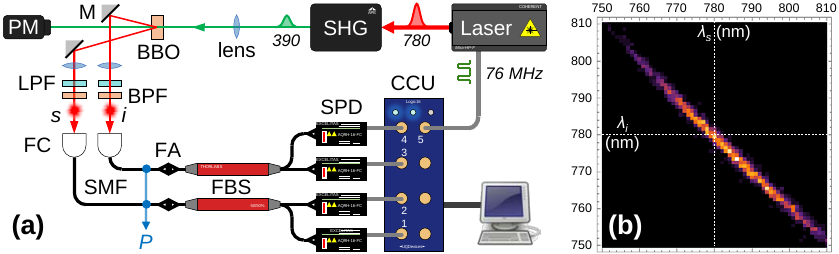}
\caption{(a) Experimental setup. SHG: second-harmonic generator; BBO: $\beta$-BaB$_2$O$_4$ crystal (thickness 1 mm); M: mirror; PM: power meter; LPF: long-pass filter; BPF: bandpass filter; FC: fiber coupler; SMF: single-mode fiber; $P$: measurand, photon number distribution of a SMF-coupled PPS; FA: fiber adaptor (mating sleeve); FBS: 50/50 fiber beam splitter; SPD: single-photon detector; CCU: coincidence counting unit. (b) Coarsely measured joint spectral intensity of the employed PPS via two tunable filters (not shown in figure, FWHM: 0.07 and 0.2~nm).}\label{fig.exp_setup}
\end{figure}

Figure~\ref{fig.exp_setup} (a) is a schematic diagram of our experimental setup. Photon pairs were generated by a type-I SPDC process of a BBO ($\beta$-BaB$_2$O$_4$ crystal, 1~mm thickness) via pulsed pump (center wavelength: 390~nm, repetition rate $r_r$: 76~MHz, pulse width: 150~fs, average power: 200~mW). After passing through long-pass filters (LPFs) that remove stray light of the pump, the joint spectral intensity (JSI) of our PPS was measured as shown in Fig.~\ref{fig.exp_setup} (b)~\footnote{The JSI plot in Fig.~\ref{fig.exp_setup} (b) is the result of coincidence counts scanned using two grating-based tunable BPFs (tBFPs) instead of FBSs after the FAs in Fig.~\ref{fig.exp_setup} (a). The FWHM of tBFPs for the $s$ and $i$ modes was 0.07 and 0.2~nm, respectively. The center wavelength of the tBPFs was moved in 1 nm steps. The reason why the JSI looks like a curve instead of a straight line is because we plotted the graphs based on wavelength, not frequency.}. We used BPFs with three different widths (FWHM, $w_1$: 1.2, $w_2$: 3, and $w_3$: 10 nm $@$ 780 nm) to estimate the $P$ matrix under various conditions. Then each photon was coupled into a SMF connected to a 50/50 fiber beam splitter (FBS) and measured by SPDs at the two outputs of the FBS. The output electronic signals from the four SPDs were counted through a CCU (time resolution: 78.125 ps). In order to accurately measure the detection probabilities, the output was received and processed together with the trigger signal from the pump laser. Time delays among all signals were adjusted, and the coincidence window was set to 2.5 ns (32 bins). Under various BPF conditions, counts for the 16 detection combinations were collected by the CCU at 1 s intervals for over 1 h. The stability of the pump power was monitored by a power meter (PM). More detailed descriptions on the experimental conditions are given in Appendix~\ref{asd}.

With the pump power fixed at around 200 mW, experiments were conducted for six cases of BPF combinations for the $s$ and $i$ ports: (i) $w_1$-$w_1$, (ii) $w_1$-$w_2$, (iii) $w_1$-$w_3$, (iv) $w_2$-$w_2$, (v) $w_2$-$w_3$, and (vi) $w_3$-$w_3$. Photon counting rates change according to the BPFs, and thus $p_g$ defined as $P_{11}$, $\eta_H$, $g^{(2)}$, and $g_h^{(2)}$ for $s$ and $i$ photons also change accordingly. In addition, the uncertainties of their estimates depend on the amount of counting data ($\propto N_m$). To investigate the tendency between uncertainties and $N_m$, a bootstrapping method (resampling with replacement)~\cite{Efron} was used. We generated 100 bootstrap samples from the original dataset of detection frequencies (empirical distribution) according to $N_m$ (sample size) for each BPF case. We estimated $P$ for all samples and then calculated $E_{rmsl}$, $p_g$, $\eta_H$, $g^{(2)}$, and $g_h^{(2)}$ from the estimated $P$. 

\begin{figure}[h]
\centering\includegraphics[width=0.482 \textwidth]{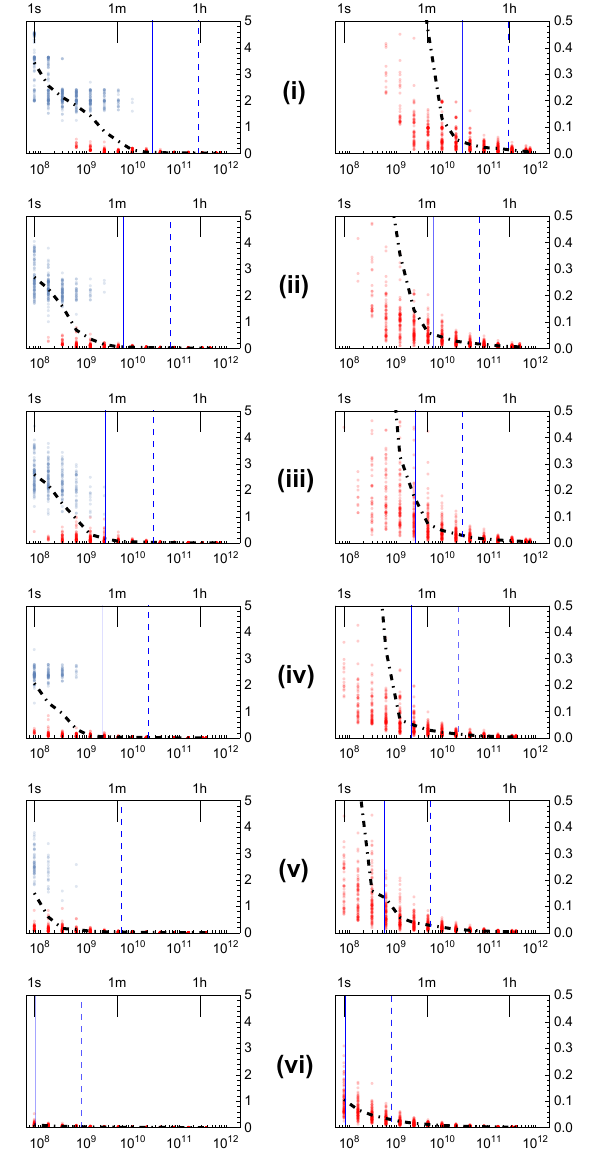}
\caption{Calculated $E_{rmsl}$ of bootstrap samples for BPF configurations of (i) to (vi). The $x$-axis is $N_m$ (equivalent to measured counts of the trigger signal, 76 MHz $\times$ $2^n$ s) expressed in $\log_{10}$ scale. The $x$-positions of the solid (dashed) lines are where the 4-fold coincidence counts of all SPDs are expected to be 10 (100). The range of $E_{rmsl}$ in the left (right) graphs is marked up to 5 (0.5) to show all results (only accurately estimated results). The dot-dashed lines are interpolation curves for the means of $E_{rmsl}$.}
\label{fig.exp_data0}
\end{figure}

Figure~\ref{fig.exp_data0} shows $E_{rmsl}$ of estimated $P$ for all samples according to $N_m$ [measured frequencies of the pump (trigger) signals collected in $2^n$ s, with the corresponding time scales marked above] for each BPF configuration. When calculating $E_{rmsl}$ of the bootstrap samples, the true $P$ in Eq.~(\ref{defRMSLE}) is unknown, so it was replaced with the estimated result using all the original data of each case. As described in Sec.~\ref{s3a}, estimation accuracy increases ($E_{rmsl}$ decreases) as $f$ (or $N_m$) increases. So for a fair comparison of the six cases, we added a vertical solid (dotted) line where the expected value of $f_{44}$ for the smallest probability $W_{44}$ is 10 (100). When $N_m$ is smaller than the solid line, the values of $E_{rmsl}$ are roughly divided into two groups (blue: $E_{rmsl}>1$, red: $E_{rmsl}\le1$). That is, when $f_{44}$ is less than 10, the estimated $P_{jk}$ (especially where $j$ or $k$ is 2) is highly erroneous, resulting in a large $E_{rmsl}$ (blue group). Conversely, when $f_{44}$ is sufficiently large ($>100$), the $E_{rmsl}$ values of all cases are less than 0.1. Therefore, the measurement frequency for the smallest detection probability (mostly the frequency of all click events, $f_{44}$) should be at least 10, while 100 or more is recommended for accurate estimation.

\begin{figure}[h]
\centering\includegraphics[width=0.75 \textwidth]{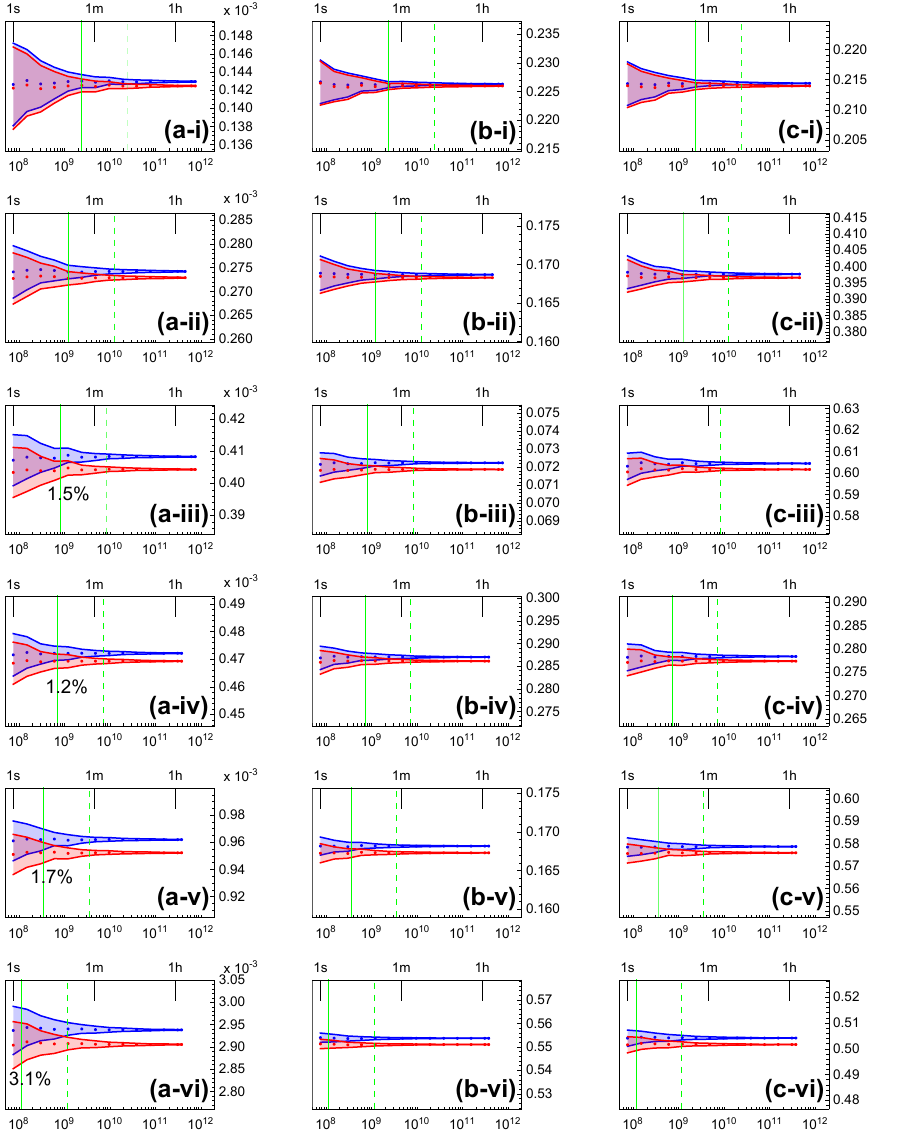}
\caption{Estimated (a) $p_g$, (b) $\eta_{H,s}$, and (c) $\eta_{H,i}$ in BPF configurations of (i) to (vi). The $x$-positions of the solid (dashed) lines are where the coincidence counts between the $s$ and $i$ ports are expected to be $10^{5(6)}$, where the relative uncertainties are similar to or less than 1 (0.4)~$\%$, except for graphs labeled with specific values.}
\label{fig.exp_data1}
\end{figure}

\begin{figure}[h]
\centering\includegraphics[width= 1 \textwidth]{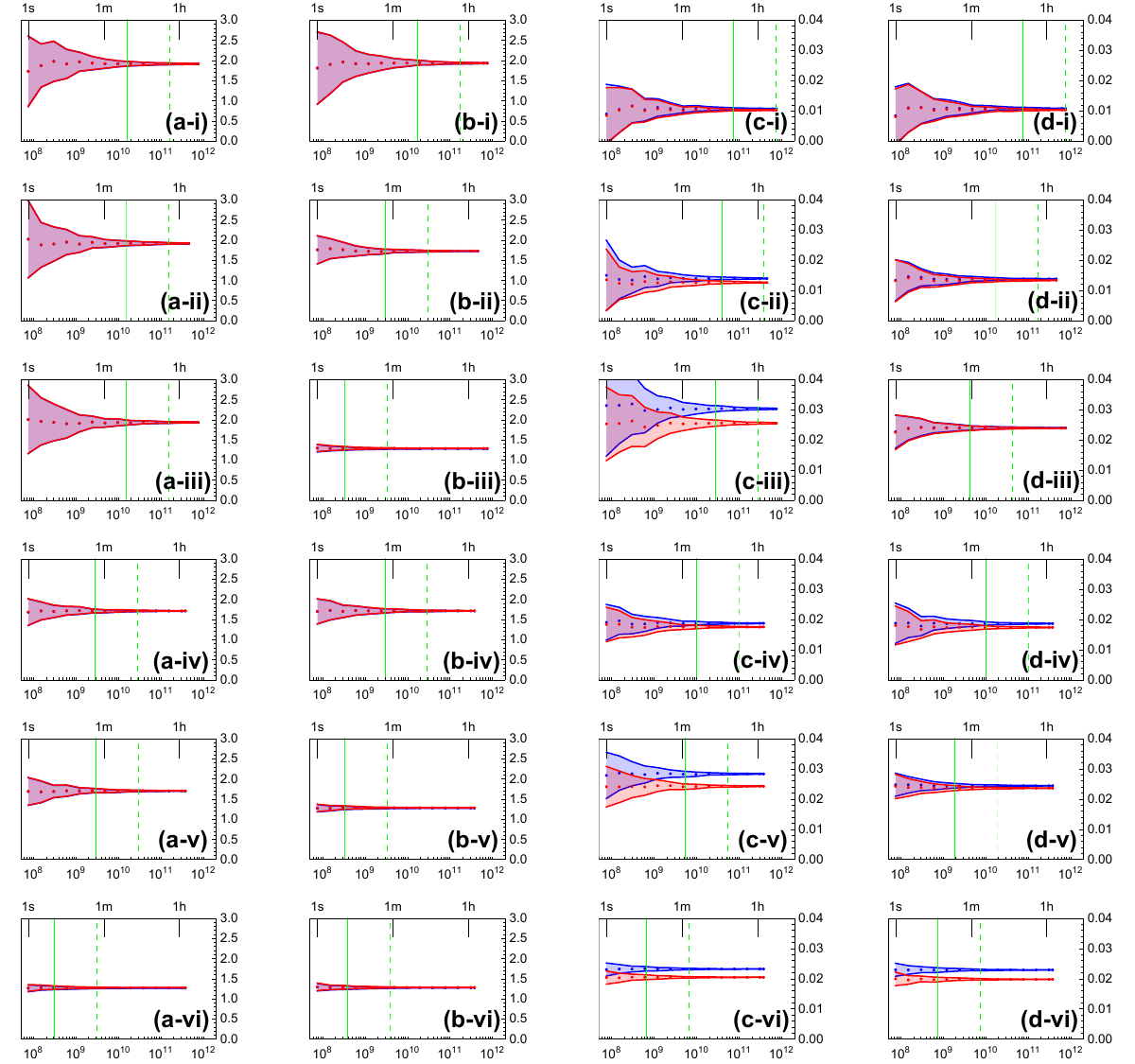}
\caption{Estimated (a, b) $g^{(2)}$ and (c, d) $g_{h}^{(2)}$ for the ($s$, $i$) modes in BPF configurations of (i) to (vi). The $x$-positions of the solid (dashed) lines in (a--d) are where the corresponding main factors of the estimates $C_{1\cap2}, C_{3\cap4}, C_{1\cap2\cap(3\cup4)}$, and $C_{(1\cup2)\cap3\cap4}$ are 10$^{3(4)}$, respectively. Uncertainties at the solid lines are about 0.1 and 0.002 for $g^{(2)}$ and  $g_h^{(2)}$, respectively.}
\label{fig.exp_data2}
\end{figure}

Since $p_g$ and $\eta_H$ mainly depend on $P^{(1)}$ while $g^{(2)}$ and $g_h^{(2)}$ are also related to $P^{(2)}$, these estimates are separately shown in Figs.~\ref{fig.exp_data1} and \ref{fig.exp_data2}. The range of the $y$-axis for all graphs in Fig.~\ref{fig.exp_data1} was fixed at 95~$\%$ to 105~$\%$ of the convergence values of the estimates with reconstructed $P$, represented in red. Therefore, all standard deviations represented as areas have the same scale in terms of the relative uncertainty, $u_r(x)=u(x)/\bar x$. For comparison, as shown in Fig.~\ref{fig.SIMg2}, the estimates from the noise-corrected counts \footnote{The estimates of $p_g$ and $\eta_H$ via (noise-corrected) counts were calculated via $p_g=\sum_{j=1}^2 \sum_{k=3}^4 \frac{C_{jk}}{\eta_{j} \eta_{k}}$ and $\eta_{H,s(i)} = {p_g} \left(\sum_{k(j)} {S_{k(j)}} ~ {\eta_{k(j)}^{-1}} \right)^{-1}$, where $C_{jk}$ is the (noise-corrected) coincidence counting rate between the $j$ and $k$ channels for the $s$ and $i$ modes, respectively, and $S_j$ and $\eta_j$ are the (noise-corrected) single counting rate and $\eta_d$ of SPD$_j$.} are shown in blue. In our experiments, the noise counts were very small (on the order of $10^{-7}$), so the estimates via raw counts are not shown because they are almost indistinguishable from the blue plots. The positions of the vertical solid (dashed) lines in the graphs correspond to the condition where the total 2-fold coincidence counts between the $s$ and $i$ modes, which is $C_{(1 \cup 2)\cap(3 \cup 4)}$ closely related to $P_{11}$, are 10$^{5(6)}$. The relative uncertainties at the solid line positions are similar to or less than 1~$\%$, except for the graphs in which specific values are indicated. As expected, the uncertainties (or standard deviations) decrease as $N_m$ increases and are small enough for 1 min of data collection. This is because $p_g$ and $\eta_H$ are determined by $P^{(1)}$, while $p_g = P_{11} \simeq O(P^{(1)}) \ge 10^{-4}$ is sufficiently larger than $1/r_r\simeq 10^{-8}$. In addition, it was observed that $p_g$ increased as the widths of the BPFs increased, and also that $\eta_{H,i}$ increased up to 60.2~$\%$ in Fig.~\ref{fig.exp_data1} (c-iii) where the difference between the widths of the BPFs for $s$ and $i$ was the largest ($w_1$-$w_3$). When the widths of the BPFs were the same, convergence of the $\eta_H$ values in Fig.~\ref{fig.exp_data1} (b, c) i, iv, and vi increased because $q_{1,2}/q_3$ decreases as the width increases. The reason why the estimates from the noise-corrected counts (in blue) are always larger than those from the reconstructed $P$ (in red) is because $p_g$ was defined as $P_{11}$ but the former (blue) were converted directly from $C_{(1 \cup 2)\cap(3 \cup 4)}$, which also contains parts of the $P_{12}$, $P_{21}$, and $P_{22}$ components. 

Unlike the $p_g$ and $\eta_H$ cases, the estimates of $g^{(2)}$ and $g_h^{(2)}$ are of similar order to each other regardless of the BPF configuration, so the range of the $y$-axis in the graphs of Fig.~\ref{fig.exp_data2} is fixed as [0, 3] and [0, 0.04], respectively. The positions of the vertical solid (dashed) lines in the graphs for \{$g_{s}^{(2)}, g_{i}^{(2)}, g_{h,s}^{(2)}, g_{h,i}^{(2)}$\} correspond to the condition where the coincidence counts of \{$C_{1\cap2}, C_{3\cap4}, C_{1\cap2\cap(3\cup4)}, C_{(1\cup2)\cap3\cap4}$\} are 10$^{3(4)}$, respectively. These coincidence counts are the main factors in the $g^{(2)}$ and $g_h^{(2)}$  estimations. The standard deviations at the solid (dashed) lines are $\le 0.13$ $(0.04)$ and $\le 0.002$ $(0.0007)$ for $g^{(2)}$ and $g_h^{(2)}$, respectively, of which the relative uncertainties are about 8 (2.6)~$\%$ or less. Compared to the results of $p_g$ and $\eta_H$, the relative uncertainties of the estimates of $g^{(2)}$ and $g_h^{(2)}$ are more than one order of magnitude higher for the same $N_m$. As discussed in Sec.~\ref{s2e3}, when the JSD has a strong correlation as shown in Fig.~\ref{fig.exp_setup} (b), $g^{(2)}$ for the $s$ or $i$ mode is determined by the width of the employed BPF for the mode. The narrower the width, the closer the effective mode number $K'$ is to 1, resulting in a $g^{(2)}$ of 2. The results in Fig.~\ref{fig.exp_data2} (a, b) demonstrate this: the convergence values of $g^{(2)}$ under BPFs corresponding to $w_1$, $w_2$, and $w_3$ were 1.92 (2), 1.71 (2), and 1.28 (2), respectively, regardless of the $s$ or $i$ mode. Furthermore, as discussed in Sec.~\ref{s3c} about Fig.~\ref{fig.SIMg2}, the estimates for $g^{(2)}$ are almost indistinguishable between blue and red, but there are differences for $g_h^{(2)}$. In particular, as shown in Fig.~\ref{fig.exp_data2} (c-iii), a smaller width of the BPF used in the heralded mode than in the heralding mode makes a large difference, since the ratio of $P_{22}$ to $P_{12}$ or $P_{21}$ is mainly dependent on the shape of the (here, fixed) JSD and the width of the BPFs.

We should note that the uncertainties in Figs.~\ref{fig.exp_data1} and~\ref{fig.exp_data2} do not include the uncertainties of other experimental parameter estimations, such as $\eta_d$ of the SPDs and the beam splitting ratios of the FBSs, but rather only include photon counting fluctuations~\footnote{That is, we assume that the values of $T(R)_{BS}$ and $\eta_d$ are fixed and there are no additional optical losses. Including the uncertainties of the estimated experimental parameters increases the final uncertainty of each characteristic. For reference, the estimations of $T(R)_{BS}$ and $\eta_d$ are discussed in Appendix~\ref{asd}.}. This is because we intended to show and analyze only the accuracy of our methodology.

\section{Conclusion}\label{s5}
In this paper, we demonstrated how the photon number distribution (or matrix) $P$ of a PPS, originally having only diagonal components, can have off-diagonal components due to spectral or spatial filtering and optical losses that occur in typical experimental situations. In particular, we detailed how to theoretically calculate $P$ from a JSD, and in the process, we also reported how to analytically calculate the effective mode number $K$. We have described how the main characteristics of PPSs, namely $p_g$, $\eta_H$, $g^{(2)}$, and $g_h^{(2)}$, are expressed by $P$. We also discussed count variation due to noise and highlighted the problems with estimating the characteristics of PPSs (especially $g_h^{(2)}$) directly from the counts, even when the noise effects on the counts are corrected. As a method was needed to accurately estimate $P$, we improved upon previous methods by utilizing all counting information with multiple SPDs. We also adopted RMSLE as an estimation accuracy metric because the previously used fidelity metric is not suitable for the $P$ of PPSs. Simulation results showed that our method offers clear improvements in terms of PND estimation accuracy, and we also reported and analyzed related experimental results for characteristics of PPSs. While we only considered up to $P^{(2)}$ and used two SPDs per mode in this paper, our method can be easily extended for the accurate estimation of $P^{(3)}$ and above via multiple SPDs.

\appendix

\section{Probability of two-pair events, $p^{(2)}$} \label{asa}
To calculate $p^{(2)}$, we first define a more generalized state $\ket \upsilon$ with normalized functions $g$ and $h$ as
\begin{align}\label{eq.upsilon}
\ket {\upsilon} 
 = & \frac{1}{2} \iint d x_1 d y_1 ~ h (x_1, y_1) ~ \hat a ^\dagger (x_1) \hat b ^\dagger  (y_1)
\iint d x_2 d y_2 ~  g(x_2, y_2) ~   \hat a ^\dagger (x_2) \hat b ^\dagger  (y_2) \ket 0 .
\end{align}
Using $\bra 0 \hat a (x_1) \hat a (x_2) \hat a ^\dagger (x_3) \hat a ^\dagger (x_4) \ket0 = \left\{\delta(x_1-x_3)\delta(x_2-x_4)+\delta(x_1-x_4)\delta(x_2-x_3)\right\}$~\cite{JMO_51_1637}, the probability of $\ket \upsilon$ is calculated as follows: 

\begin{align}
4 P(\ket \upsilon) = & ~ 4 \braket {\upsilon} {\upsilon} \notag \\  
= & \bra 0 \int \cdots \int d x_1 \cdots d x_4 ~ d y_1 \cdots d y_4~ h^* (x_1, y_1) g^* (x_2, y_2) h (x_3, y_3) g (x_4, y_4) \notag \\
& ~~~~~~~~~~~~~~~~~~~~ \times \hat a (x_1) \hat a (x_2) \hat a ^\dagger (x_3) \hat a ^\dagger (x_4)
\hat b  (y_1)  \hat b  (y_2)  \hat b ^\dagger  (y_3)  \hat b ^\dagger  (y_4) \ket 0 \notag \\
= & \int \cdots \int d x_1 \cdots d x_4 ~ d y_1 \cdots d y_4~ h^* (x_1, y_1) g^* (x_2, y_2) h (x_3, y_3) g (x_4, y_4) \notag \\
& ~~~~~~~~~~~~~~~~~~~~ \times \left\{\delta(x_1-x_3)\delta(x_2-x_4)+\delta(x_1-x_4)\delta(x_2-x_3)\right\} \notag \\
& ~~~~~~~~~~~~~~~~~~~~ \times \left\{\delta(y_1-y_3)\delta(y_2-y_4)+\delta(y_1-y_4)\delta(y_2-y_3)\right\} \notag \\
= & ~~~~ \iiiint dx_1 dx_2 dy_1 dy_2 ~ h^\ast (x_1,y_1) g^\ast (x_2,y_2) h(x_1,y_1) g(x_2,y_2) \notag \\
& + \iiiint dx_1 dx_2 dy_1 dy_2  ~ h^\ast (x_1,y_1) g^\ast (x_2,y_2) h(x_1,y_2) g(x_2,y_1) \notag \\
& + \iiiint dx_1 dx_2 dy_1 dy_2  ~ h^\ast (x_1,y_1) g^\ast (x_2,y_2) h(x_2,y_1) g(x_1,y_2) \notag \\
& + \iiiint dx_1 dx_2 dy_1 dy_2  ~ h^\ast (x_1,y_1) g^\ast (x_2,y_2) h(x_2,y_2) g(x_1,y_1) \label{eq.upsilon_prob1} \\
= & \iint dx_1 dy_1 ~ \abs {h (x_1,y_1)}^2  \iint dx_2 dy_2 ~ \abs {g (x_2,y_2)}^2  \notag \\ 
&  + O_y  + O_x  + \abs {\iint dx_1 dy_1 h^\ast(x_1,y_1) g(x_1,y_1)}^2  \notag \\
= & 1 + O_y  + O_x  + O, \label{eq.upsilon_prob2}
\end{align} 
where $O$ is the overlap between the $g$ and $h$ functions and $O_{x(y)}$ is the partial overlap on the $x(y)$-axis as defined in 3rd (2nd) term of Eq.~(\ref{eq.upsilon_prob1}). The physical meaning of the above formula is essentially similar to (or an extended version of) the probability of  $\frac{1}{2} \left(1+ \left| \int d\omega f_1^\ast(\omega) f_2(\omega) \right|^2 \right)$ for the state $\frac{1}{\sqrt 2}\hat{a}_{f_1} ^\dagger \hat{a}_{f_2} ^\dagger \ket 0$ defined using $\hat{a}_{f_j} ^\dagger = \int d \omega f_j(\omega) \hat{a} ^\dagger (\omega)$~\cite{Barnett}. First, let us assume that $g$ and $h$ are two-dimensional Gaussian functions with very small widths. If the $x$ and $y$ centers of the two functions are very far apart, then four independent photons are generated, specifically $\ket \upsilon=\frac{1}{2}\hat a^\dagger \hat b^\dagger \hat c^\dagger \hat d^\dagger \ket 0 =\frac{1}{2} \ket{1,1,1,1}$, so that $P(\ket \upsilon)=1/4$. On the other hand, if the $x~(y)$ centers are the same but the $y~(x)$ centers are far apart, then $\ket \upsilon$ is the same as $\frac{1}{2} \hat a^\dagger {}^2 \hat b^\dagger \hat c^\dagger \ket 0 = \frac{1}{\sqrt2} \ket{2,1,1}$ and $P(\ket \upsilon)=1/2$. In the case of $g=h$, $\ket \upsilon=\frac{1}{2}\hat a^\dagger {}^2 \hat b^\dagger {}^2 \ket 0 = \ket{2,2}$ and $P(\ket \upsilon)=1$. Therefore, according to the comprehensive overlap of the two functions, $P(\ket \upsilon)$ has a value in $[1/4,1]$. Second, for the purposes of our calculations, consider a more general case. If $g$ and $h$ functions are described as $\sum_j v_j A_j(x)B_j(y)$ and $\sum_k w_k C_k(x)D_k(y)$ in multiple modes, then $O_{x}$ and $O_{y}$ are represented as overlaps of only the functions corresponding to the $x$ and $y$ variables expressed as $\sum_{j,k} ~ | v_j|^2  | w_k|^2 ~ \left| \int dx A_j^*(x) C_k(x) \right|^2$ and $\sum_{j,k} ~ | v_j|^2  | w_k|^2 ~ \left| \int dy ~ B_j^*(y) D_k(y) \right|^2$, respectively. In the case of $g=h$, $P(\ket \upsilon)$ is simplified as $\frac{1}{2}\left( 1 +  \sum_j |v_j|^4 \right)=\frac{1}{2}(1+1/K)$. In other words, even if $g$ and $h$ are the same, $P(\ket \upsilon)$ varies in [1/2, 1] depending on the effective mode number of the function, since the generation of the second photon pair (by $h$) in a different mode (from $g$) lowers the overall generation probability from 1. Finally, the approximated probability of two-pair events $p^{(2)}$ in the absence of BPFs is expressed by the effective mode number $K$ of $f(\omega_s,\omega_i)$ as
$$
p^{(2)}  \simeq  |\xi|^4 \frac{1}{2}\left(1+\frac{1}{K}\right).
$$

Additionally, the analytic calculation of $1/K$ is as below:
\begin{align} \label{eq.KK}
& \iiiint dx_1 dx_2 dy_1 dy_2  ~ f^\ast (x_1,y_1) f^\ast (x_2,y_2) f(x_1,y_2) f(x_2,y_1) \notag \\
= & \iiiint dx_1 dx_2 dy_1 dy_2  ~\sum_i c_i^\ast A_i^\ast(x_1) B_i^\ast(y_1) ~ \sum_j c_j^\ast A_j^\ast(x_2) B_j^\ast(y_2)\notag\\
& ~~~~~~~~~~~~~~~~~~~~~~~~~~~~~~~~~~\times \sum_k c_k A_k(x_1) B_k(y_2) ~\sum_l c_l A_l(x_2) B_l(y_1) \notag \\
= & \sum_{ijkl} c_i^\ast c_j^\ast c_k c_l \int dx_1 A_i^\ast(x_1) A_k(x_1) \int dx_2 A_j^\ast(x_2) A_l(x_2) \notag \\
& ~~~~~~~~~~~~~~~~~~~~~~\times \int dy_1 B_i^\ast(y_1) B_l(y_1) \int dy_2 B_j^\ast(y_2) B_k(y_2) \notag \\
= & \sum_{ijkl} c_i^\ast c_j^\ast c_k c_l \delta_{ik} \delta_{jl} \delta_{il} \delta_{jk} =  \sum_{i} |c_i|^4 = 1/K.
\end{align}

\section{Probability distribution $P^{(2)}$ from $p^{(2)}$ via spectral filtering} \label{asb}
There are various types of BPFs such as Gaussian and flat-top depending on the purpose and design, but none are ideal (distorted shape and $T_{max}<1$). The spectral transmittance $T(\omega)$ of a BPF is relatively easy to measure and can be modeled as a spectral-dependent beam splitter~\cite{NJP_12_063001}. Thus, the creation operators ($\hat a_s^\dagger$, $\hat a_i^\dagger$) in Eq.~(\ref{eq.expansion2-2}) are replaced by superposed ones of transmitted ($\hat a^\dagger$, $\hat b^\dagger$) and reflected ($\hat c^\dagger$, $\hat d^\dagger$) modes, 
\begin{align}
\hat a_s ^\dagger (\omega_s) & \rightarrow t_x(x) \hat a ^\dagger (x) + r_x(x) \hat c ^\dagger (x) \\
\hat a_i ^\dagger (\omega_i) & \rightarrow t_y(y) \hat b ^\dagger (y) + r_y(y) \hat d ^\dagger (y).
\end{align}
For a simple and intuitive notation, instead of $s$ and $i$ subscripts, we classify the creation operators alphabetically according to the modes and mark $x~(y)$ instead of $\omega_{s(i)}$. The co-efficiencies $t_{z}(z)$ and $r_{z}(z)$ are spectral amplitudes of the transmission and reflection (for $z=x,y$) of a BFP, respectively, and satisfy the normalization condition as $|t_{z}(z)|^2+|r_{z}(z)|^2=1$. Then the JSD is decomposed as
$$f(x,y)= \sqrt{q_1} F_1(x,y) + \sqrt{q_2} F_2(x,y)+\sqrt{q_3} F_3(x,y)+\sqrt{q_4} F_4(x,y)$$
with normalized functions $\{F_j\}$ 
\begin{align*}
F_1(x,y) & = t_x(x) r_y(y)f(x,y)/\sqrt{q_1} \\
F_2(x,y) & = r_x(x) t_y(y)f(x,y)/\sqrt{q_2} \\
F_3(x,y) & = t_x(x) t_y(y)f(x,y)/\sqrt{q_3} \\
F_4(x,y) & = r_x(x) r_y(y)f(x,y)/\sqrt{q_4}, 
\end{align*}
where $\{q_j\}$ are normalization factors (or probabilities) for $\iint dx dy |F_j(x,y)|^2=1$. Then $\ket{\psi_1}$ in Eq.~(\ref{eq.expansion1}) is represented as
\begin{align*}
\ket {\psi_1} 
 = \xi \iint d x_1 d y_1 ~  &
\left\{ \sqrt{q_1} F_1 (x_1, y_1) ~ \hat a ^\dagger (x_1) \hat d ^\dagger  (y_1) 
+\sqrt{q_2} F_2 (x_1, y_1) ~ \hat c ^\dagger (x_1) \hat b ^\dagger  (y_1) \right. \\
& \left. 
+ \sqrt{q_3} F_3 (x_1, y_1) ~ \hat a ^\dagger (x_1) \hat b ^\dagger  (y_1)  
+ \sqrt{q_4} F_4 (x_1, y_1) ~ \hat c ^\dagger (x_1) \hat d ^\dagger  (y_1) \right\} \ket 0.
\end{align*}
The four cases of $F_j$, ($TR$, $RT$, $TT$, $RR$)$_{si}$, for single photon-pair events are distinguished from each other via their creation operators ($\hat a ^\dagger \hat d ^\dagger$, $\hat c ^\dagger \hat b ^\dagger$, $\hat a ^\dagger \hat b ^\dagger$, $\hat c ^\dagger \hat d ^\dagger$), so that the number distribution of transmitted photons from $p^{(1)}$ is expressed by $\{q_j\}$ the same as $P^{(1)}$ in Eq.~(\ref{eq.P(1)matrix}).

With spectral filtering, the state of two photon-pair events in Eq.~(\ref{eq.expansion2-2}) is rewritten as
\begin{align*}
\ket {\psi_{2'}} 
 = \frac{\xi^2}{2} \iint d x_1 d y_1 ~  &
\left\{ \sqrt{q_1} F_1 (x_1, y_1) ~ \hat a ^\dagger (x_1) \hat d ^\dagger  (y_1) 
+\sqrt{q_2} F_2 (x_1, y_1) ~ \hat c ^\dagger (x_1) \hat b ^\dagger  (y_1) \right. \\
& \left. 
+ \sqrt{q_3} F_3 (x_1, y_1) ~ \hat a ^\dagger (x_1) \hat b ^\dagger  (y_1)  
+ \sqrt{q_4} F_4 (x_1, y_1) ~ \hat c ^\dagger (x_1) \hat d ^\dagger  (y_1) \right\} \\ 
 \iint d x_2 d y_2 ~  &
\left\{\sqrt{q_1} F_1 (x_2, y_2) ~ \hat a ^\dagger (x_2) \hat d ^\dagger  (y_2) 
 +\sqrt{q_2} F_2 (x_2, y_2) ~ \hat c ^\dagger (x_2) \hat b ^\dagger  (y_2) \right.\\
& + \left. \sqrt{q_3} F_3 (x_2, y_2) ~ \hat a ^\dagger (x_2) \hat b ^\dagger  (y_2) 
 + \sqrt{q_4} F_4 (x_2, y_2) ~ \hat c ^\dagger (x_2) \hat d ^\dagger  (y_2) \right\}
\ket 0 .
\end{align*}
When we calculate $\braket{\psi_{2'}}{\psi_{2'}}$, there are $4^4$ terms of 
$$\bra0 \int \cdots \int dx_1 \cdots dy_4 ~ F_j^* (x_1,y_1) F_k^* (x_2,y_2) F_l (x_3,y_3) F_m(x_4,y_4) ~ \hat x (x_1) \cdots \hat y ^\dagger (y_4) ~ \ket0,$$
where $\hat x^\dagger$ and $\hat y^\dagger$ represent ($\hat a^\dagger$ or $\hat c^\dagger$) and ($\hat b^\dagger$ or $\hat d^\dagger$) for $s$ and $i$ photons, respectively. However, due to the canonical commutation relation, there are only 36 nonzero probabilities. These can be classified into three categories using the indices $\{j,k,l,m\}$ of $F$ functions as follows: 
\begin{enumerate}
\item 4 cases of $j=k=l=m$;
\item 24 cases of $(j,k)=(l,m)$ for $j\neq k$, i.e., 6 cases ($jk$ = 12, 13, 14, 23, 24, 34) times 4 subsets $(jkjk, jkkj, kjjk, kjkj)$;
\item 8 remaining cases where all operators occur once, such as $\hat a \hat b \hat c \hat d \hat a^\dagger  \hat b^\dagger \hat c^\dagger \hat d^\dagger$, i.e., (1234, 1243, 2134, 2143, 3412, 3421, 4312, 4321).
\end{enumerate}
Calculations similar to Eq.~(\ref{eq.upsilon_prob2}) are performed, and the following conclusions are obtained for each category. 
\begin{enumerate}
\item The four cases ($j=1,2,3,4$) correspond to $P^{(2)}_{20}$, $P^{(2)}_{02}$, $P^{(2)}_{22}$, $P^{(2)}_{00}$, respectively, and each probability is described as $|\xi|^4 q_j^2 (1+1/\kappa_j)/2$ in the form of Eq.~(\ref{eq.prob2}), where $\kappa_j$ is the effective mode number of $F_j$.
\item For the case of $jk=12$ or $34$, ($F_j$, $F_k$) is associated with completely different mode operators, such as ($\hat a \hat d$, $\hat c \hat b$) or ($\hat a \hat b$, $\hat c \hat d$), respectively, so that the total probability (including all four subsets) is only $|\xi|^4 q_j q_k$ and contributes to $P^{(2)}_{11}$. For the cases of $jk=13$ and $24$, there are two $\hat a^\dagger$ and two $\hat c^\dagger$ for $s$ photons, so that the partial overlap $O_x^{(jk)}=O_x(F_j,F_k)$ is added to the basic probability of $|\xi|^4 q_j q_k$ and this contributes to $P^{(2)}_{21}$ and $P^{(2)}_{01}$, respectively. In the cases of $jk=14$ and $23$, similarly $O_y^{(jk)}=O_y(F_j,F_k)$ is added and contributes to $P^{(2)}_{10}$ and $P^{(2)}_{12}$, respectively.
\item All cases are related to a complex overlap value (or its complex conjugate) defined as $$O_c = \iiiint dx_1 \cdots dy_2 ~ F_1^* (x_1,y_1) F_2^* (x_2,y_2) F_3 (x_1,y_2) F_4(x_2,y_1),$$ so that the total probability is $2 |\xi|^4 \sqrt{q_1 q_2 q_3 q_4} ~ {\rm Re}[O_c]$ and contributes to $P^{(2)}_{11}$.
\end{enumerate}
Finally, the exact form of $P^{(2)}$ is given by
\begin{gather}\label{eq.P(2)matrix}
P^{(2)} = |\xi|^4\left(
\begin{tabular} {ccc}
$q_4^2 ~ (1+1/\kappa_4)/2$     & $q_2 q_4 \left( 1+ O_x^{(24)}\right)$               &$q_2^2 ~ (1+1/\kappa_2)/2$\\
$q_1 q_4 \left( 1+ O_y^{(14)} \right)$   & \begin{tabular}{c} $q_1 q_2 + q_3 q_4$\\ $+ 2 \sqrt{q_1 q_2 q_3 q_4}~ {\rm Re}[O_c]$  \end{tabular}    &$q_2 q_3 \left( 1+ O_y^{(23)} \right)$\\
$q_1^2 ~ (1+1/\kappa_1)/2$     & $q_1 q_3 \left( 1+ O_x^{(13)} \right)$               &$q_3^2~ (1+1/\kappa_3)/2$
\end{tabular} \right).
\end{gather}
We note that when the JSD is separable and BFPs are ideal, all $\kappa_j$ and all overlaps in Eq.~(\ref{eq.P(2)matrix}) are unity and $q_1 q_2 = q_3 q_4$ is satisfied, so that Eq.~(\ref{eq.P(2)matrix}) becomes Eq.~(\ref{eq.P(2)matrix_O}), which is also equivalent to the result of the simple loss model.

\section{Effect of spatial filtering on a PPS} \label{asc}
The state of a single photon pair for spatially correlated PPSs can be described similarly as in Appendix~\ref{asa} as
\begin{align*}
\ket {\psi_1} \simeq & ~\xi_{s} \iint d\vec r_1  d\vec r_2 ~ f(\vec r_1, \vec r_2) ~ \hat a_s ^\dagger (\vec r_1)  \hat a_i ^\dagger (\vec r_2) \ket 0\\
= &~ \xi_{s}\iint d\vec r_1  d\vec r_2 
\left(
\sqrt{q_1}F_1(\vec r_1, \vec r_2) \hat a ^\dagger (\vec r_1)  \hat d ^\dagger (\vec r_2) 
+\sqrt{q_2}F_2(\vec r_1, \vec r_2) \hat c ^\dagger (\vec r_1)  \hat b ^\dagger (\vec r_2) \right.\\
& ~~~~~~~~~~~~~~~~~~~~~~
\left.+~ \sqrt{q_3}F_3(\vec r_1, \vec r_2) \hat a ^\dagger (\vec r_1)  \hat b ^\dagger (\vec r_2)
+\sqrt{q_4}F_4(\vec r_1, \vec r_2) \hat c ^\dagger (\vec r_1)  \hat d ^\dagger (\vec r_2)
\right)
\ket 0 ,
\end{align*}
where the spatial correlation is expressed as $f(\vec r_1, \vec r_2)$ and ${F_j}$ is defined as
\begin{align*}
F_1(\vec r_1, \vec r_2) & = f(\vec r_1, \vec r_2) ~ m_s(\vec r_1) r_i(\vec r_2) / \sqrt{q_1}\\
F_2(\vec r_1, \vec r_2) & = f(\vec r_1, \vec r_2) ~ r_s(\vec r_1) m_i(\vec r_2) / \sqrt{q_2}\\
F_3(\vec r_1, \vec r_2) & = f(\vec r_1, \vec r_2) ~ m_s(\vec r_1) m_i(\vec r_2) / \sqrt{q_3}\\
F_4(\vec r_1, \vec r_2) & = f(\vec r_1, \vec r_2) ~ r_s(\vec r_1) r_i(\vec r_2) / \sqrt{q_4}.
\end{align*}
Spatial filtering is replaced by a position-dependent beam splitter via
\begin{align*}
\hat a_s ^\dagger (\vec r_1) & \rightarrow m_s(\vec r_1) \hat a ^\dagger (\vec r_1)+ r_s(\vec r_1) \hat c ^\dagger (\vec r_1)\\
\hat a_i ^\dagger (\vec r_2) & \rightarrow m_i(\vec r_2) \hat b ^\dagger (\vec r_2) + r_i(\vec r_2) \hat d ^\dagger (\vec r_2),
\end{align*}
where $m_{s(i)}$ is the spatial mode of the aperture or SMF used for $s~(i)$ photons, and $r_{s(i)}$ is the residual mode for normalization. This is the same expression as that for spectral filtering; the only difference is that the integration space is doubled.

\section{Experimental parameters} \label{asd}

\subsection{$T$ and $R$ of FBSs} 

\begin{figure}[h]
\centering\includegraphics[width=0.4\textwidth]{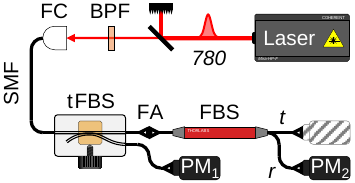}
\caption{Experimental setup for measuring $T$ and $R$ of a FBS. tFBS: tunable FBS (tunable directional coupler).}\label{exp_BS}
\end{figure}

The transmittance $T$ and reflectance $R$ of the two FBSs in Fig.~\ref{fig.exp_setup} (a) were evaluated with the setup shown in Fig.~\ref{exp_BS}. We first attenuated the intensity of the pulsed pump laser (center wavelength: 780 nm), narrowed the linewidth with a BPF of $w_2$, and split with a SMF-based tunable FBS (tFBS, tunable directional coupler), whose output intensities were around 20 (at PM$_1$) and 200 $\mu$W. One output of the tFBS was measured with PM$_1$ to monitor the intensity of the input light to the FBS, and the other output was connected to the input of the FBS. Then the two outputs of the FBS were measured with PM$_2$. When we denote the measured intensity at PM$_j$ as $I_j$ and the intensity ratio ($I_2 / I_1$) for the $t~(r)$ mode of the FBS as ${\mathcal I}_{t(r)}$, then $T~(R)$ of the FBS is calculated as ${\mathcal I}_{t(r)}/({\mathcal I}_t+{\mathcal I}_r)$. We measured ${\mathcal I}_{t(r)}$ 200 times to obtain its standard uncertainty and calculated the standard uncertainty of $T~(R)$ of the FBSs via error propagation. The estimates of $T$ were 0.4952 (3) and 0.4846 (3) for the two FBSs used in our experiments. Since we originally assumed lossless BSs, $R$ can be calculated as $1-T$ and has the same uncertainty.

\subsection{$\eta_d$ and $d$ of SPDs}

\begin{figure}[h]
\centering\includegraphics[width=0.41\textwidth]{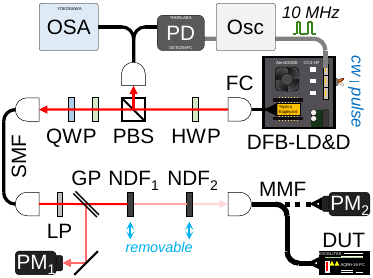}
\caption{Experimental setup for measuring $\eta_d$ of SPDs. DFB-LD\&D: distributed-feedback laser diode (Toptica Eagleyard, 780 nm) and driver (AeroDIODE, CCS-HP), PBS: polarizing beam splitter, H(Q)WP: half (quarter) wave plate, OSA: optical spectrum analyzer, PD: photodetector (Thorlabs, DET025AFC), Osc: oscilloscope, LP: linear polarizer, GP: glass plate (1.5 mm), NDF: neutral density filter, MMF: multi-mode fiber, DUT: device under test (SPD).}\label{exp_SPD}
\end{figure}

Our method for estimating $P$ requires information about all experimental parameters. The most important parameter is $\eta_d$ of the SPDs used in experiments. Accurate (low-uncertainty) measurement of $\eta_d$ is a topic under research by national metrology institutes (NMIs). Our experiments for $\eta_d$ estimation were performed in consultation with the Photometry \& Radiometry Group (PRG) at the Korea Research Institute of Standards and Science (KRISS). The analysis of the measurement uncertainty of $\eta_d$ is beyond the scope of this paper, but it is roughly expected to be greater than $5 \%$. For reference, standard uncertainties for $\eta_d$ performed by PRGs of NMIs are less than 1 \%~\cite{JMO_62_1732, M_57_015002}. Our experimental setup shown in Fig.~\ref{exp_SPD} is similar to the setup of \cite{JMO_62_1732}, where $\eta_d$ is estimated via the photon counting rate of the attenuated laser source detected by the SPD. This is because the total number of photons per second is determined by the laser power, and the detected counting rate is reduced by $\eta_d$. The experimental sequence is as follows. Transmittance $T_j$ of two neutral density filters (NDF$_j$) is first estimated using PM$_1$ and PM$_2$. Then with both NDFs installed, $\eta_d$ is estimated from the measured photon counting rate. At this time, the laser intensity incident on the SPD can be adjusted via a half wave plate before the polarizing beam splitter and monitored with PM$_1$. Our experimental setup differs from that of \cite{JMO_62_1732} by roughly three factors: (1) the laser source was not power stabilized and was driven in cw or pulsed ($<$ 2 ns, 10 MHz) mode; (2) we did not precisely recalibrate the photodiode sensors of the power meters; (3) $\eta_{d}$ was measured for a SPD coupled to a multi-mode fiber (50 micron MMF). The first factor, laser in cw or pulsed mode, did not significantly affect the $\eta_d$ estimate within the expected uncertainty ($>$ 5 \%), consistent with the results of \cite{M_57_015002}. In addition, power stabilization is an assumption for Poisson distribution, which is unlikely to have a significant impact because Poisson and thermal distributions do not significantly differ when the average photon number is very small ($\bar n \ll 1$). The second factor significantly increased the uncertainty of $\eta_d$ estimation. As described in \cite{JMO_62_1732}, in general the uncertainty of $\eta_d$ is mainly from $T$ of NDFs and the photodiodes. In this experiment, the uncertainties of the photodiodes also affected the uncertainty of $T$, so the total uncertainty of $\eta_d$ in our experiments is expected to be much larger than the manufacturer's specification of 5 \% (Thorlabs, S122C). The reason for the third factor (using a MMF-coupled SPD) is measurement tolerance. The presence or absence of NDFs can cause a slight tilt of the laser beam, which is too sensitive for SMF coupling and can cause additional loss. In addition, the active area of the SPD used is 180~$\mu$m, which is larger than the core size of the MMF, so there is no additional loss. Even within the active area, the response varies depending on location, but we ignored this. Due to the dead time effect discussed in \cite{JMO_62_1732}, we measured \{$\eta_j$\} ($\eta_d$ of SPD$_j$) by setting the counting rates of the SPDs to less than $10^5$ cps, and the estimates are \{56.2, 57.5, 56.7, 54.8\} \%.

Since $\{ d_j \}$ of SPDs include not only dark counts of the SPDs but also noise counts due to stray light, these should be measured in actual experimental setups like that shown in Fig.~\ref{fig.exp_setup}~(a) when the pump beam is blocked. The dark count rate of the employed SPDs (Excelitas, SPCM-AQRH-16-FC) is around 11 cps, but due to stray light, the count rates go up to tens of cps. However, $\{ d_j \}$ could be reduced by a factor of $0.19=2.5/13.15$ by coincidence counting with the trigger signal received from the pump laser, where the coincidence window and the period of the trigger signal were 2.5 ns and 1/(76 MHz) $\simeq$~13.15 ns, respectively. From 1000 data of counts for 1 s, $\{ d_j \}$ were estimated as \{1.01(1), 2.11(2), 0.94(1), 1.00(1)\} $\times$ 10$^{-7}$.

\section*{Funding}
This work was supported by Korea Research Institute of Standards and Science (KRISS) projects (GP2024-0013-04, -0012-17), Institute of Information \& communications Technology Planning \& Evaluation (IITP) grant (2021-0-00185), and the National Research Council of Science \& Technology (NST) grant (CAP22051-000) funded by the Korea government (MSIT). 

\section*{Acknowledgments}
We thank Dr. Dong-Hoon Lee (KRISS) for an informative discussion on the $\eta_d$ estimation of single photon detectors.

\bibliography{peta}

\end{document}